\documentclass[final,5p,times,twocolumn,sort&compress]{elsarticle}
\usepackage[T1]{fontenc}
\usepackage[utf8]{inputenc}
\usepackage{amsmath,amssymb,bm}
\usepackage{graphicx,booktabs,array,longtable,adjustbox}
\usepackage{placeins,flafter}
\usepackage[unicode,hidelinks]{hyperref}
\journal{Measurement}
\usepackage{stfloats,flushend}
\begin{document}
\begin{frontmatter}
\title{Inverse identification of thermophysical parameters for reduced effective-temperature uncertainty in an ion optical clock}
\author[a,e]{Jie Zhao}
\author[a,b,c,d]{Yani Zuo\corref{cor1}}
\ead{zuoyn@nim.ac.cn}
\author[a,b,c,d]{Shaoyang Dai}
\author[a,f]{Chaowei Wang}
\author[g]{Ang Huang}
\author[g]{Jian Song}
\author[g]{Guangyao Wang}
\author[h]{Fangfei Wu}
\author[i]{Haotian Shi}
\author[e]{Zhengyong Li\corref{cor1}}
\ead{zhyli@bjtu.edu.cn}
\author[a,b,c,d]{Fang Fang\corref{cor1}}
\ead{fangf@nim.ac.cn}
\affiliation[a]{organization={Time and Frequency Research Institute, National Institute of Metrology},city={Beijing},postcode={100029},country={China}}
\affiliation[b]{organization={Hefei National Laboratory},city={Hefei},postcode={230088},country={China}}
\affiliation[c]{organization={Key Laboratory of State Administration for Market Regulation (Time Frequency and Gravity Primary Standard)},city={Beijing},postcode={100029},country={China}}
\affiliation[d]{organization={Beijing Key Laboratory of Quantum Metrology Technology and Instruments, National Institute of Metrology},city={Beijing},postcode={100029},country={China}}
\affiliation[e]{organization={Department of Physics, School of Physical Science and Engineering, Beijing Jiaotong University},city={Beijing},postcode={100044},country={China}}
\affiliation[f]{organization={Department of Precision Machinery and Precision Instruments, School of Engineering Science, University of Science and Technology of China},city={Hefei},state={Anhui},postcode={230022},country={China}}
\affiliation[g]{organization={Division of Thermophysics Metrology, National Institute of Metrology},city={Beijing},postcode={100029},country={China}}
\affiliation[h]{organization={College of Sciences, China Jiliang University},city={Hangzhou},postcode={310018},country={China}}
\affiliation[i]{organization={College of Metrology Measurement and Instrument, China Jiliang University},city={Hangzhou},postcode={310018},country={China}}
\cortext[cor1]{Corresponding authors.}
\begin{abstract}
The blackbody-radiation (BBR) shift remains a leading systematic uncertainty of room-temperature single-ion optical clocks. Accurate evaluation of this shift requires the effective temperature at the ion position, which is difficult because RF heating, low and spatially varying surface emissivity and inaccessible trap surfaces make this quantity difficult to measure directly. Here, we developed a hybrid inversion framework that combines radiance-domain infrared thermography, black-painted high-emissivity reference regions, a finite-element model of a \textsuperscript{171}Yb\textsuperscript{+} end-cap ion trap and a reduced-order surrogate for repeated forward calculations and parameter identification. The surrogate reproduced the FE test fields with a full-field mean absolute error (MAE) of 0.02220 mK. The calibrated model predicted the ion-position effective temperature with a combined standard uncertainty of 155.0 mK, corresponding to a temperature-related BBR frequency-shift uncertainty of 1.647 \ensuremath{\times} 10\textsuperscript{-19} for the \textsuperscript{2}S\textsubscript{1/2}–\textsuperscript{2}F\textsubscript{7/2} (467 nm) E3 transition. This method provides a reproducible route for reducing and tracing the BBR uncertainty of room-temperature trapped-ion optical clocks.
\end{abstract}

\end{frontmatter}
\section{Introduction}\label{main-sec-1}
Optical clocks based on trapped ions and neutral atoms have undergone a period of transformative progress\cite{m01,m02}, with state-of-the-art systems reaching systematic uncertainties at the 10\textsuperscript{-19} level \cite{m03,m04,m05,m06,m07,m08,m09}, nearly three orders of magnitude beyond present cesium fountain primary standards. This places them at the centre of the planned redefinition of the SI second \cite{m10} and enables local time-scale generation \cite{m11}, relativistic geodesy and chronometric levelling \cite{m12}, millimetre-scale gravitational-redshift measurements \cite{m15}, searches for ultralight dark matter using atomic clocks and optical cavities \cite{m13}, and entangled optical-clock networks \cite{m14}.

Among the systematic effects of room-temperature clocks, the blackbody radiation (BBR) shift is typically the largest single contributor to the uncertainty budget \cite{m16,m17}. Suppressing this shift and establishing a traceable uncertainty for its evaluation are prerequisites for operation at systematic uncertainties below 10\textsuperscript{-18}. For trapped atoms and ions, the BBR shift can be expressed as \cite{m18,m19}:

\begin{equation}
\Delta\nu_{\mathrm{BBR}}=-\frac{\Delta\alpha_0}{2h}\left\langle E^2(T)\right\rangle[1+\eta(T)]
\label{main-eq-1}
\end{equation}
where \ensuremath{\Delta}\textit{\ensuremath{\alpha}}\textsubscript{0} is the differential static polarizability of the clock transition, \textit{\ensuremath{\eta}} (\textit{T}) is a small dynamic correction, h is Planck's constant and $\left\langle E^2(T)\right\rangle=[831.9430(15)\,\mathrm{V}/\mathrm{m}]^2(T/300\,\mathrm{K})^4$ is the mean-squared BBR electric field at temperature \textit{T} \cite{m30}. For \textsuperscript{171}Yb\textsuperscript{+} \textsuperscript{2}S\textsubscript{1/2}-\textsuperscript{2}F\textsubscript{7/2} (E3) clock transition, the atomic-response parameters are taken consistently from the mid-infrared polarizability-transfer data set of Ref. \cite{m33}. The corresponding spectral transfer from the mid-infrared reference wavelength to the BBR spectrum is described in Supplementary Sections \hyperref[supplement-sec-S3.2]{S3.2}–\hyperref[supplement-sec-S3.4]{S3.4}. These quantities describe the atomic response to BBR, whereas $\langle E^2\rangle$ is determined by the local thermal-radiation environment and is therefore characterized here by the apparatus-specific effective temperature T\textsubscript{eff}.

For a radio-frequency ion trap, T\textsubscript{eff} is neither directly measurable nor safely assumed equal to the chamber temperature. RF dissipation in the feedthrough, spacers and electrodes heats the trap structure, producing a non-uniform temperature distribution over the surfaces seen by the ion \cite{m17,m61,m62,m31}. In the trap studied here, the resulting ion-position effective-temperature rise is approximately 0.43 K, showing that RF-induced trap heating cannot be neglected at the 10\textsuperscript{-19} level.

\begin{figure*}[!htbp]
\centering
\hypertarget{main-fig-1}{}
\includegraphics[width=170.0000mm,keepaspectratio]{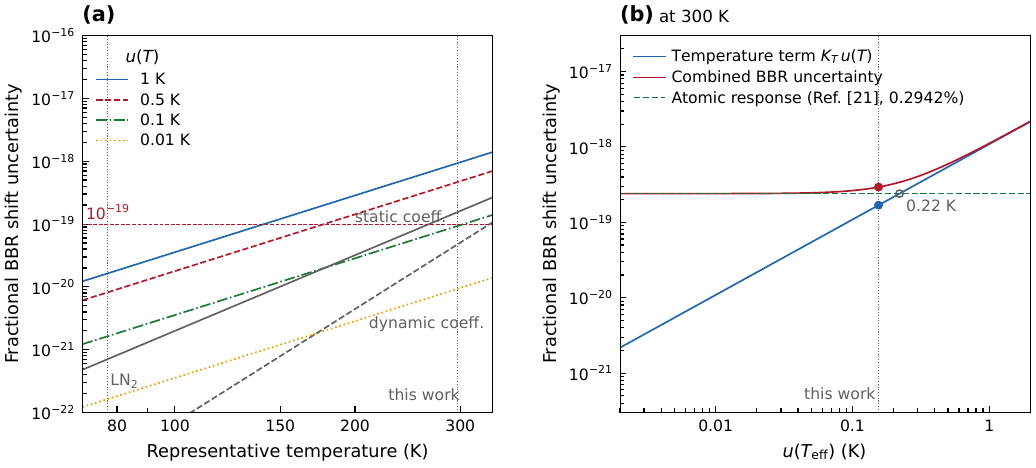}
\caption{Fractional BBR frequency shift uncertainty of the \textsuperscript{171}Yb+ (E3) clock transition. (a) Fractional BBR frequency shift uncertainty versus representative temperature, for temperature uncertainties of 1, 0.5, 0.1, and 0.01 K. Solid and dashed grey lines show the contributions from the static and dynamic polarizability coefficients, respectively, calculated using the atomic-response parameters reported in \cite{m33}; the vertical dotted lines mark 77 K (LN\textsubscript{2}) and the room temperature operating point used in this work. (b) Fractional BBR shift uncertainty at 300 K as a function of the temperature uncertainty $u(T_{\mathrm{eff}})$. The blue curve shows the temperature-related contribution, the green dashed line shows the atomic-response contribution corresponding to a relative standard uncertainty of 0.2942\%, and the red curve shows their quadrature sum. The open marker at $u(T_{\mathrm{eff}})=0.2206\,\mathrm{K}$indicates where the temperature-related and atomic-response contributions are equal. The vertical dotted line marks the combined standard uncertainty $u(T_{\mathrm{eff}})=0.1550\,\mathrm{K}$obtained in this work.}\label{fig1}
\end{figure*}

The required accuracy of T\textsubscript{eff} follows from the atomic-response coefficients in Eq.~\eqref{main-eq-1}. For the \textsuperscript{171}Yb\textsuperscript{+} E3 transition \cite{m32,m35}, these atomic-response parameters contribute a fractional BBR uncertainty of 2.40\ensuremath{\times}10\textsuperscript{-19} at 300 K \cite{m33}. As illustrated in Fig. \ref{fig1}, the temperature-induced BBR uncertainty becomes comparable to the atomic-response contribution for a temperature uncertainty of approximately 0.2206 K. Reducing the temperature uncertainty to approximately 0.10 K lowers the temperature-related contribution to the 10\textsuperscript{-19} lever. Below this range, the combined BBR uncertainty is increasingly limited by the atomic-response coefficients rather than by the temperature evaluation.

Available atomic-response determinations do not fully agree within their quoted uncertainties. Central values and uncertainties are therefore taken from a single internally consistent data set rather than mixed between different determinations; the alternative data sets and the selection adopted here are discussed in Supplementary Sections \hyperref[supplement-sec-S3.3]{S3.3} and \hyperref[supplement-sec-S3.4]{S3.4}.

Several strategies have been developed to suppress the BBR shift or reduce the uncertainty in its evaluation. At the apparatus level, cryogenic operation can strongly suppress the BBR shift and its uncertainty. Engineered thermal-radiation shields have also been proposed for SI-traceable thermometry based on optical lattice clocks \cite{m26}. Such approaches have achieved a BBR Stark-shift uncertainty of 1.7\ensuremath{\times}10\textsuperscript{-20} in an optical lattice clock \cite{m28} and a total systematic uncertainty of 4.4\ensuremath{\times}10\textsuperscript{-19} in a liquid-nitrogen-cooled \textsuperscript{40}Ca\textsuperscript{+} ion clock \cite{m04}. These approaches, however, require dedicated thermal design and are not directly applicable to an existing room-temperature trap. The BBR shift can also be determined from clock-frequency changes induced by controlled changes in the thermal environment, including RF-induced trap heating \cite{m33,m34}; this requires frequency comparisons with sufficient averaging. In parallel, improved atomic-response data \cite{m20,m21,m22,m23} and clock transitions with low BBR sensitivity \cite{m09,m21} reduce the atomic contribution, but do not eliminate the need to determine T\textsubscript{ef}\textsubscript{f}. For room-temperature neutral-atom clocks, T\textsubscript{eff} can be evaluated using calibrated in-vacuum temperature probes combined with finite-element (FE) modelling, reaching uncertainties at the level of tens of millikelvin \cite{m24,m25,m27}. In ion clocks, this route is largely closed, since direct contact thermometry near the trap perturbs the RF field. The effective temperature is therefore inferred from accessible-point sensors and infrared thermography combined with FE simulation \cite{m17,m62,m31}, with uncertainties as low as 42 mK recently reported \cite{m05}. Characterization of the BBR field can dominate an ion-clock uncertainty budget \cite{m29}. For an existing room-temperature ion trap, accurate evaluation of T\textsubscript{eff} therefore relies critically on the fidelity of the thermal model and its experimental constraints. 

The accuracy of an FE-based T\textsubscript{eff} evaluation is limited by the thermophysical parameters supplied to the model. Surface emissivities, dielectric loss tangents, thermal-contact properties, and boundary temperatures are often imperfectly known for an assembled ion trap. In particular, metallic emissivities depend on surface morphology, oxidation, and temperature itself \cite{m36,m37,m38}, while dielectric loss and thermal-contact parameters may vary between nominally identical components. Treating these quantities only through literature values or broad prior intervals therefore propagates directly into the uncertainty of the RF-induced temperature field and ultimately into T\textsubscript{eff}.

Infrared thermography provides substantially more information than is normally used in such evaluations. A thermogram contains spatially resolved radiance over many thousands of pixels and several materials, yet in previous ion-trap thermal evaluations, thermography has been used primarily to identify heated regions or to compare experiment and simulation at selected locations. The spatial radiance field has not generally been exploited as a direct inverse constraint for identifying uncertain thermophysical parameters in the FE model. Consequently, discrepancies between experiment and simulation do not necessarily reveal which thermophysical parameter is responsible, since similar deviations can originate from multiple sources and may require additional measurements to distinguish them \cite{m17}.

Surrogate-assisted estimation, physics-informed parameter identification, and multi-condition FE model updating have been investigated in related measurement inverse problems \cite{m40,m41,m42}. Inverse heat conduction has also been used to estimate heat sources and internal temperatures \cite{m43}. In this work, the uncertain thermophysical parameters are identified directly from spatially resolved infrared radiance measurements. A coupled electromagnetic-thermal FE model of a \textsuperscript{171}Yb\textsuperscript{+} end-cap trap supplies the RF-induced temperature-rise field, and a sensitivity-enriched reduced-order surrogate, motivated by non-intrusive neural reduced-order modelling and derivative-informed projection \cite{m44,m45}, replaces the expensive FE solve during iterative identification while preserving the weak parameter-sensitivity directions the inversion depends on. Parameter identification is performed in the radiance domain, where emissivity has a more direct influence on the measured signal than on the temperature field, helping to reduce its coupling with ambient temperature during inversion. Calibrated black-painted regions provide emissivity-fixed radiometric anchors that further break the degeneracy. The retrieved parameters are reintegrated into the full FE model to evaluate T\textsubscript{eff} and its uncertainty. The identified model is further tested at a second RF operating condition.

The resulting model gives an ion-position effective-temperature rise of 0.4345 K and a combined standard uncertainty of 0.1550 K. This corresponds to a temperature-related fractional BBR uncertainty of 1.647 x 10\textsuperscript{-19}. Combined with the atomic-response contribution of 2.33\ensuremath{\times}10\textsuperscript{-19}, the total fractional BBR uncertainty is 2.850\ensuremath{\times}10\textsuperscript{-19}, with the temperature evaluation accounting for approximately 33.40\% of the total variance. The corresponding RF-induced change in the fractional BBR shift is 4.607\ensuremath{\times}10\textsuperscript{-19}. The method therefore provides a traceable route for evaluating T\textsubscript{eff} in room-temperature ion trap.

\section{Measurement framework and radiometric constraints}\label{main-sec-2}
\subsection{Measurand and overall measurement strategy}\label{main-sec-2.1}
The measurand of this work is the effective blackbody-radiation temperature at the ion position, T\textsubscript{eff}, which determines the thermal-radiation field entering the BBR Stark shift in Eq.~\eqref{main-eq-1}. For the non-uniform radiative environment of an ion trap, T\textsubscript{eff} can be defined as the temperature of an isotropic blackbody radiation field that has the same radiative energy density at the ion position as the actual trap environment. Following Doležal et al. \cite{m17}, it is evaluated in the FE model from the equilibrium temperature of a virtual, unity-emissivity blackbody probe placed at the ion position. The numerical implementation of this probe is described in Section \ref{main-sec-3.2}.

\begin{figure*}[!htbp]
\centering
\hypertarget{main-fig-2}{}
\includegraphics[width=145.0000mm,height=0.70\textheight,keepaspectratio]{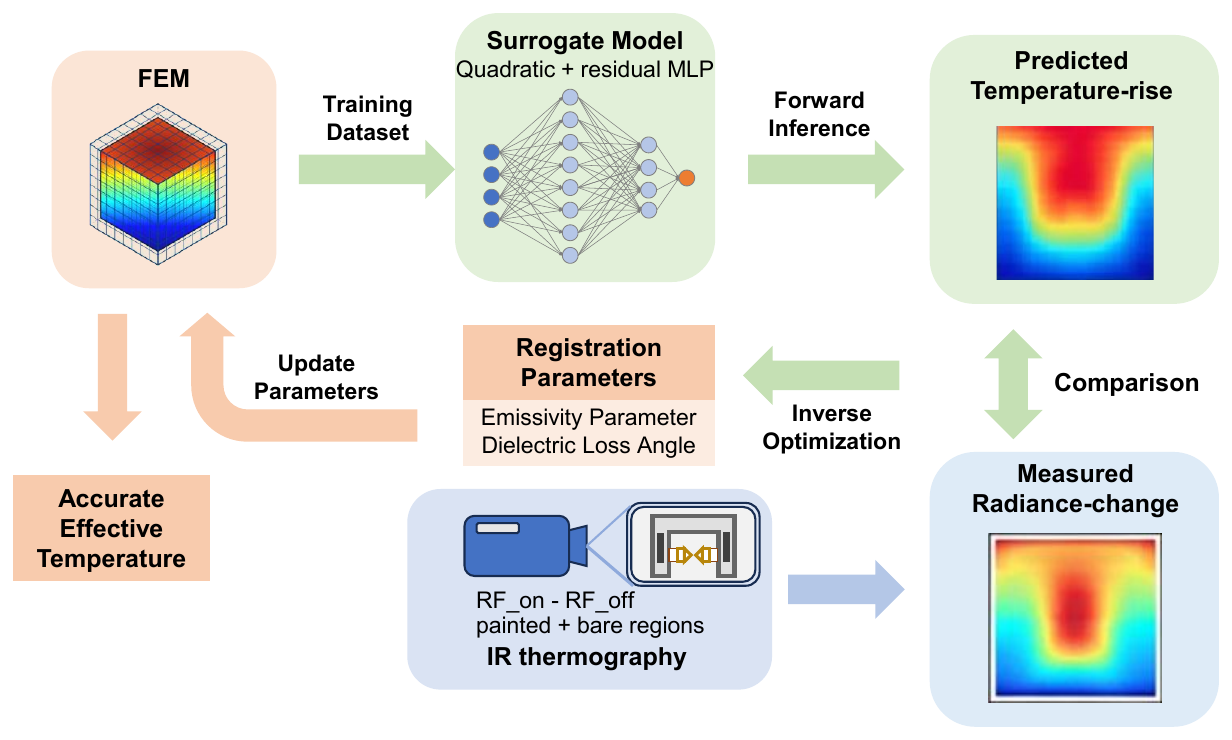}
\caption{Workflow of the proposed BBR effective-temperature evaluation method. Calibrated infrared radiance measurements, with the black-painted region providing a fixed high-emissivity reference, are used to constrain surrogate-assisted thermophysical parameter identification. The retrieved parameters are then reintroduced into the full FE model to evaluate the effective temperature at the ion position, followed by uncertainty propagation.}\label{fig2}
\end{figure*}

Because most surfaces contributing to the ion's radiative environment are not directly accessible to thermography, T\textsubscript{eff} cannot be determined from the infrared images alone. In the present method, calibrated RF-induced radiance changes measured over the visible trap surfaces are used to identify the uncertain thermophysical parameters through a surrogate-assisted inverse procedure. The identified parameters are subsequently returned to the full FE model to calculate T\textsubscript{eff}. The overall measurement and evaluation sequence is summarized in Fig. \ref{fig2}.

\subsection{Ion-trap system, RF operating conditions and reference thermometry}\label{main-sec-2.2}
\begin{figure}[!htbp]
\centering
\hypertarget{main-fig-3}{}
\includegraphics[width=85.0000mm,height=0.70\textheight,keepaspectratio]{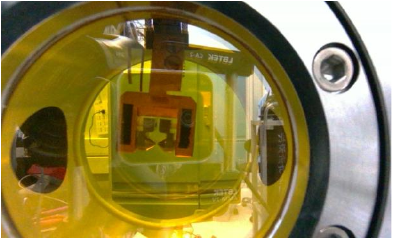}
\caption{Structure of the \textsuperscript{171}Yb\textsuperscript{+} end-cap ion trap, showing the Mo end-cap, Mo RF electrodes, C-shaped Cu carrier alumina insulating bolts, fused-silica spacers, and positioning rings.}\label{fig3}
\end{figure}

The actual structure of the end-cap ytterbium ion trap employed in our system is shown in Fig. \ref{fig3}. The trap contains RF electrodes and end-cap electrodes, all made from molybdenum (Mo) with gold-sputter surfaces. Both the feedthrough ceramics and the insulating bolts are made of alumina. The alumina bolts pass through fused-silica spacers positioned between the C-shaped Copper (Cu) carrier and the end-cap electrodes, while fused-silica positioning rings are placed between the C-shaped Cu carrier and the Mo RF electrodes. The stainless-steel enclosure surrounding the trap is evacuated and acts as the thermal boundary and heat sink. To enable infrared thermometry using the infrared (IR) camera, the auxiliary chamber is equipped with ZnSe viewports, while the peak wavelength of room-temperature thermal radiation is approximately 10 \ensuremath{\mu}m, well matched to the camera's detection band. 

All measurements reported here were taken at an RF frequency of 20.408 MHz. Thermograms were acquired in three states: RF-off, which provides the baseline for the radiance subtraction; a 500 V condition used for thermophysical parameter identification; and a 485 V condition used for cross-condition validation. 

Contact thermometry served both as the reference for the radiometric calibration of Section \ref{main-sec-2.3.2} and as the ambient boundary input to the FE model. During the infrared measurements, two calibrated Pt100 probes were positioned on the left and right sides of the ZnSe viewport to monitor the ambient temperature around the trap. The measured ambient temperature was supplied as a known operating-condition input to the subsequent surrogate calculations and parameter inversion. Prior to installation, the ten Pt100 resistance thermometers were calibrated at the Thermal Metrology Institute of the National Institute of Metrology (NIM) against a Standard Platinum Resistance Thermometer (SPRT) traceable to the ITS-90. The Pt100 sensors and the SPRT were simultaneously immersed in a precision temperature bath (TL-1010S) at four setpoints (291.15, 296.15, 301.15 and 306.15 K), spanning the temperature range relevant to the trap measurements.

At each setpoint, readings were acquired over 1000 s and averaged to reduce random temperature fluctuations. The standard deviations of the SPRT-derived temperature during the averaging period were 0.020, 0.051, 0.054 and 0.158 mK at the four setpoints, respectively, confirming the temporal stability of the bath. For each Pt100 channel, a linear calibration relation $T_{\mathrm{true}}=a\cdot T_{\mathrm{display}}+b$ was fitted against the four SPRT reference temperatures, yielding a maximum fitting residual of 2.25 mK across all ten channels (channel 4).

The uncertainty budget of the calibration is summarized in Table \ref{tab1}. The two largest contributions were the curve-fitting residual (2.02 mK) and the SPRT's traceable calibration uncertainty (2.00 mK). Along with the bath's spatial inhomogeneity (1.00 mK) and its 1000 s temporal stability (0.16 mK), the quadrature sum of the four factors gives a combined calibration uncertainty of 3.02 mK for the worst-case channel.

\begin{table}[!htbp]
\centering
\caption{~Calibration error sources of the Pt100 resistance thermometers using an ITS-90-traceable SPRT.}\label{tab1}
{\small\setlength{\tabcolsep}{3pt}
\begin{adjustbox}{max width=\linewidth}
\begin{tabular}{@{}lr@{}}
\toprule
Uncertainty source & Uncertainty \\
\midrule
Curve fitting & 2.02 mK \\
SPRT calibration uncertainty & 2.00 mK \\
Bath inhomogeneity & 1.00 mK \\
Bath stability & 0.16 mK \\
\midrule
Total & 3.02 mK \\
\bottomrule
\end{tabular}
\end{adjustbox}}
\end{table}
\subsection{Broadband infrared thermography with black-painted references}\label{main-sec-2.3}
To obtain reliable temperature information at the trap surfaces, spatially resolved broadband radiance was combined with black-painted reference regions of calibrated emissivity to constrain the radiance-domain inversion described in Chapter \ref{main-sec-4}. All infrared measurements used a calibrated FLIR T1040 infrared (IR) camera with a nominal spectral response range of 7–14 \ensuremath{\mu}m and a 1024 \ensuremath{\times} 768-pixel focal-plane array.

\subsubsection{Design and role of the reference regions}\label{main-sec-2.3.1}
Black paint was applied to the C-shaped Cu structure, which lies outside the ion's direct radiative field of view and was chosen to minimize perturbation to the RF field. The black-paint emissivity was calibrated to be 0.82.

The black-painted region and metallic regions were processed in the radiance domain. The RF-on and RF-off thermograms were converted from raw DN values into calibrated broadband radiance and subtracted to obtain the RF-induced radiance change,$\Delta L$, which was compared directly with the predicted radiance change during inverse identification.

Although the black-paint emissivity is known, the black-painted region was retained in the radiance domain rather than converted separately to temperature. Such a conversion would require the absolute radiance level and a baseline surface temperature, whereas the RF-on/RF-off radiance difference cancels the common calibration offset and can be compared directly with the radiometric forward model. The residuals from all regions can therefore be evaluated consistently in the same measurement domain.

Because the black-painted region has a fixed and relatively high emissivity, its radiance change is more directly sensitive to the RF-induced surface temperature rise and less affected by reflected background radiation than that of the bare metallic surfaces. It therefore provides an emissivity-fixed radiometric constraint on the amplitude of the predicted temperature-rise field, helping to reduce the degeneracy between the heating-related parameters and the unknown metallic emissivities.

\subsubsection{Calibration of the FLIR T1040–window system}\label{main-sec-2.3.2}
\begin{figure}[!htbp]
\centering
\hypertarget{main-fig-4}{}
\includegraphics[width=65.0000mm,height=0.70\textheight,keepaspectratio]{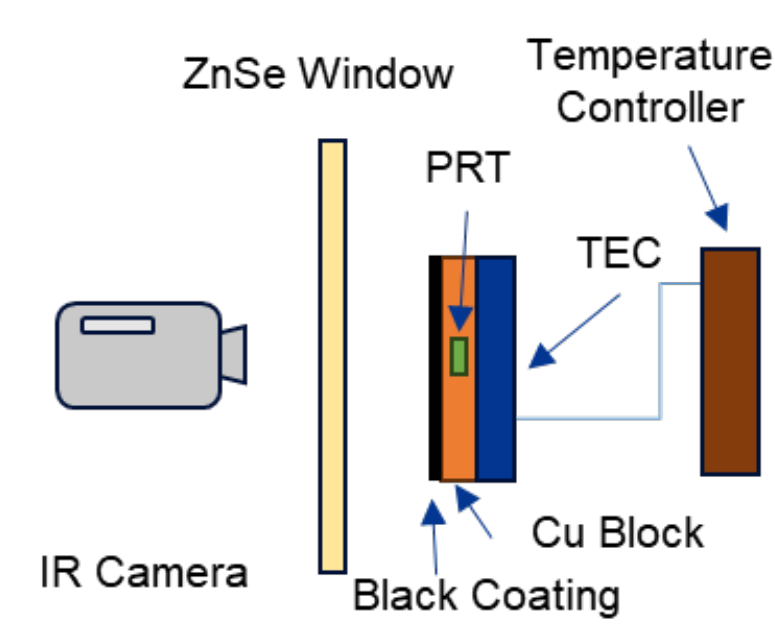}
\caption{~Radiometric calibration setup of the FLIR T1040–ZnSe window measurement path, including the IR camera, ZnSe window, temperature-controlled reference blackbody, and calibrated Pt100 probe.}\label{fig4}
\end{figure}

A temperature-controlled reference blackbody source with an effective emissivity of 0.997 was employed as the reference source. The calibration was conducted with a ZnSe window sample identical to that installed on the auxiliary chamber placed between the blackbody source and the IR camera so that the window transmission was included in the calibration. The schematic of the calibration setup of the experimental system is presented in Fig. \ref{fig4}.

The calibration procedure was conducted over a temperature range of 294.14–312.13 K, covering the possible temperature range of the ion trap. With the integration time and imaging settings of the IR camera fixed, raw 16-bit thermograms were acquired at ten temperature points separated by approximately 2 K. Before data acquisition, a non-uniformity correction (NUC) was performed manually once. The automatic NUC function was then disabled throughout the subsequent acquisition to prevent shutter-induced changes in the pixel offsets and camera signal baseline. At each temperature point, seven consecutive frames were recorded at intervals of approximately 10 s over a total duration of approximately 60 s. For each acquired thermogram, a 5\ensuremath{\times}5-pixel region of the image position corresponding to the embedded Pt100 probe was selected, and the digital number (DN) values within this region of interest (ROI) were extracted. The DN values were then averaged spatially over the ROI and temporally over the seven frames to obtain the response at each temperature point. The band-integrated blackbody radiance $L$  over the nominal 7–14 \ensuremath{\mu}m spectral response range of the IR camera was calculated from the Pt100 reference temperature using the Planck law and the effective emissivity of the reference source. This procedure thus incorporated both the ZnSe window's transmission and the camera's radiometric response into the calibration.

\begin{figure}[!htbp]
\centering
\hypertarget{main-fig-5}{}
\includegraphics[width=85.0000mm,keepaspectratio]{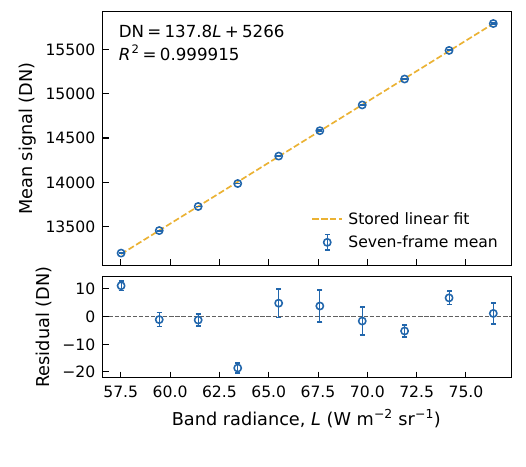}
\caption{~Radiometric calibration of the FLIR T1040–ZnSe window measurement path. The upper panel shows the DN–radiance (L) linear fit, and the lower panel shows the corresponding residuals with standard-error bars.}\label{fig5}
\end{figure}

Subsequently, a linear regression model was applied to fit the measured DN values against the calculated band-integrated radiance:

$\mathit{DN}=G\cdot L+B$where $G$ and $B$ denote the responsivity gain and offset of the camera–window system, respectively. As shown in Fig. \ref{fig5}, the measured DN values exhibit a highly linear dependence on the calculated radiance, yielding

\begin{equation}
\mathit{DN}=137.8\cdot L+5266.3
\label{main-eq-18}
\end{equation}
where $R^2=0.9999$. The fitted calibration function was subsequently used to convert the raw trap thermograms into calibrated broadband radiance for RF-on/RF-off subtraction and radiance-domain parameter inversion.

The calibration uncertainty near 300 K was evaluated from four contributions, listed in Table \ref{tab2}. The residual standard error of the DN–radiance fit was 8.58 DN. At 300 K, this corresponded to a relative radiance uncertainty of 0.0982\%, equivalent to a calibration temperature uncertainty of 61.2 mK. Temporal repeatability, taken as the standard error of the seven frame-wise ROI means at the 300.13 K setpoint, was 1.90 DN (13.5 mK); local spatial dispersion, taken as the standard deviation of the 25 frame-averaged pixel values, was 1.09 DN (7.79 mK); and the calibrated Pt100 reference probe contributed 3.01 mK. Treated as independent, their quadrature sum gave 63.2 mK at 300.13 K.

\begin{table}[!htbp]
\centering
\caption{~Uncertainty budget for the radiometric calibration of the FLIR T1040–window system at 300 K.}\label{tab2}
{\small\setlength{\tabcolsep}{3pt}
\begin{adjustbox}{max width=\linewidth}
\begin{tabular}{@{}lr@{}}
\toprule
Uncertainty source & Uncertainty \\
\midrule
Regression residual (DN--radiance linear fit) & 61.2 mK \\
Temporal repeatability & 13.5 mK \\
Spatial dispersion & 7.79 mK \\
Pt100 reference probe & 3.01 mK \\
Combined standard uncertainty & 63.2 mK \\
\bottomrule
\end{tabular}
\end{adjustbox}}
\end{table}
\subsubsection{Thermogram acquisition, registration, and preprocessing}\label{main-sec-2.3.3}
Broadband infrared images of the ion trap were acquired through the ZnSe viewport under both RF-on and RF-off conditions. To suppress reflected ambient thermal radiation, the external optical path was shielded with a high-absorptivity black flocking cloth. Camera position, viewing angle, focus and integration settings were unchanged throughout the measurement. For each operating state, seven consecutive raw 16-bit frames were recorded and averaged pixel by pixel to reduce temporal detector noise. The ambient temperature was monitored simultaneously by two calibrated Pt100 probes described in Section \ref{main-sec-2.2}. The measured ambient temperature was used as a known operating-condition input for the subsequent surrogate calculations and parameter inversion, while the difference between the two probe readings was used to bound how well a single scalar temperature represents the radiative environment of the trap and was included in the ambient-boundary uncertainty analysis of Section \ref{main-sec-6.4}.

The averaged raw images were converted into calibrated radiance using the relation obtained in Section \ref{main-sec-2.3.2},

\begin{equation}
L=(\mathit{DN}-5266.32)/137.7723.
\label{main-eq-22}
\end{equation}
Feature-based image registration can use robust consensus fitting to reject mismatched correspondences \cite{m56}. Here, before subtraction, the averaged RF-on and RF-off images were spatially registered using subpixel phase cross-correlation of their Gaussian-gradient representations. This representation suppresses the nearly uniform radiance change between the two states while preserving spatial features relevant to the registration, allowing the relative translation to be estimated from the structural information in the two images. Because the two images were acquired with the same camera geometry, only a small relative translation was corrected. The radiance change induced by RF heating was then calculated as

\begin{equation}
\Delta L(\mathbf{r})=\frac{\overline{\mathit{DN}}_{\mathrm{on}}(\mathbf{r})-\overline{\mathit{DN}}_{\mathrm{off}}(\mathbf{r})}{137.7723}
\label{main-eq-23}
\end{equation}
where the overbar denotes the average over the seven acquired frames. Because the offset term in the calibration relation is common to the two measurements, it cancels directly in the RF-on/off subtraction.

After registration, material masks were defined on the radiance-change map for the visible copper, gold, and black-painted surfaces, and were used to extract the corresponding material-specific regions for subsequent inversion.

Although the RF-on/off subtraction largely suppresses the static reflected background, changes in reflected radiance between the two operating states cannot be completely eliminated. This residual contribution is particularly relevant for the low-emissivity metallic surfaces, whose measured radiance contains both surface emission and reflected background radiation. These residual reflection effects were therefore treated explicitly in the radiometric forward model described in Section \ref{main-sec-4.2}. The resulting radiance distributions over the Cu, Au cap-plane region (Au-cap), Au end-plane region (Au-end), and black-painted region constitute the primary experimental constraints for the parameter identification in Section \ref{main-sec-4}.

\section{Coupled physical and surrogate models}\label{main-sec-3}
This chapter establishes the coupled physical model and reduced-order surrogate used for effective-temperature evaluation and thermophysical parameter identification. The coupled electromagnetic–thermal finite-element (FE) model of Sections \ref{main-sec-3.1} and \ref{main-sec-3.2} maps the surface emissivities, dielectric loss tangents, operating conditions, and thermal boundary inputs to the RF-induced temperature-rise field. The final T\textsubscript{eff} is obtained using the full FE model. A single FE calculation requires approximately 10 min, whereas parameter identification requires repeated forward evaluation. Therefore, a surrogate trained on a database of FE solutions (Sections \ref{main-sec-3.3} and \ref{main-sec-3.4}) is used during the iterative identification process. Because inverse identification depends not only on the predicted temperature field but also on its parameter sensitivities, the surrogate is verified for both predictive accuracy and sensitivity fidelity (Section \ref{main-sec-3.5}). FE+surrogate

\subsection{Electromagnetic–thermal finite-element model}\label{main-sec-3.1}
Previous methods mostly utilized an FE model to compute the geometric effective solid angle alone, while relying on discrete point measurements for the component temperatures. In an operational ion trap, however, the radio-frequency (RF) field induces localized thermal gradients near the trap electrodes and insulating ceramics through Joule heating and dielectric losses, and these variations cannot be captured by macroscopic spatial averaging. The RF field was therefore incorporated into the FE model.

The geometric model was established to the actual chamber and imported into FE simulation. The constructed FE model is presented in Fig. \ref{fig6}(a). The feedthrough was integrated into the model to capture the non-negligible heat generated by the ceramic insulation, which is conducted to the trap. The stainless-steel enclosure was defined as a vacuum domain, serving as a heat sink.

The model was developed in two steps. 

First, to evaluate the heat generation, an RF field was excited using the Electromagnetic Waves interface. At the operating frequency of 20.408 MHz, the small skin depth of the metallic conductors, approximately 14 \ensuremath{\mu}m in copper, makes the surface-current distribution important for evaluating ohmic losses, while dielectric losses depend on the local electric-field distribution within the insulating components. The Impedance Boundary Condition was applied to the Cu and Mo surfaces to calculate the Joule heating, whereas a loss-tangent model was assigned to the alumina and fused-silica components to account for dielectric losses. 

Bidirectional electro-thermal coupling was implemented through a linear temperature-dependent resistivity model and solved iteratively until a self-consistent steady state was reached.

\subsection{Thermal boundary conditions, radiative-transfer, and FE-model verification}\label{main-sec-3.2}
The second part of calculation was the heat dissipation by the Heat Transfer in Solids and Surface-to-Surface Radiation interfaces, with the power-dissipation distribution obtained in the first step serving as the heat source. Four boundary conditions were used to close the problem: the radiative exchange inside the evacuated enclosure, the thermal contact conductance at the bolted joints, convection from the external surfaces to the laboratory air, and a fixed temperature where the trap is conductively tied to the helical resonator.

Inside the evacuated chamber, radiative exchange between the trap components and the stainless-steel enclosure was modeled using a diffuse-grey surface-to-surface radiation model. The corresponding surface emissivities were assigned according to the material properties and subsequently treated as either identified or fixed parameters in the radiometric analysis.

The interfaces between assembled trap components were treated as ideal thermal contacts in the nominal FE model. To evaluate the effect of finite contact conductance, the contact coefficient was varied from 2\ensuremath{\times}10\textsuperscript{2} to 1\ensuremath{\times}10\textsuperscript{4}~W\,m\textsuperscript{-2}\,K\textsuperscript{-1}. A nominal value of 1\ensuremath{\times}10\textsuperscript{3}~W\,m\textsuperscript{-2}\,K\textsuperscript{-1} gives a T\textsubscript{eff} correction of \ensuremath{-}14.2 mK, and the associated uncertainty is evaluated in Section \ref{main-sec-6.5}.

For the surfaces exposed to the laboratory environment outside the vacuum, including the external surfaces of the feedthrough and stainless-steel chamber, a natural convective heat-transfer coefficient of 5 W\,m\textsuperscript{-2}\,K\textsuperscript{-1} was applied as a nominal model boundary assumption. Standard heat-transfer treatments \cite{m59} and vertical-plate free-convection correlations \cite{m58} provide background for such estimates; their predictions depend on geometry and flow conditions and do not independently validate this coefficient for the assembled apparatus. To account for heat conduction from the helical resonator to the feedthrough, the temperature at the top of the copper rod was measured as 298.6 K using the infrared camera and high-emissivity (0.997) black tape, calibrated against the reference blackbody source described in Section \ref{main-sec-2.3.2}, thereby avoiding direct-contact sensing that could perturb the RF field. 

To evaluate the BBR effective temperature seen by the ion, an 80 \ensuremath{\mu}m-diameter blackbody sphere with unity emissivity and a thermal conductivity of 1000 W m\textsuperscript{-1} K\textsuperscript{-1} was placed at the ion position to represent the ion \cite{m17}. Its steady-state temperature was taken as T\textsubscript{eff}, and Ref. \cite{m17} reported variations within approximately $\pm$0.01 K over the tested sphere sizes.

The FE calculations in this section used the reference thermophysical parameters summarized in Table \ref{tab5}, together with an RF frequency of 20.408 MHz, an RF amplitude of 500 V, and an ambient temperature of 296.69 K. The converged surface temperature field was then used as the training target for the surrogate model in Section \ref{main-sec-3.3}.

\begin{figure*}[!htbp]
\centering
\hypertarget{main-fig-6}{}
\begin{minipage}{175mm}
\begin{minipage}[b]{0.30695\linewidth}\centering \includegraphics[width=\linewidth,height=0.70\textheight,keepaspectratio]{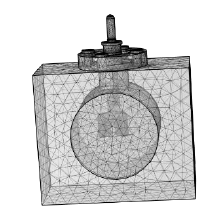}\par (a)\end{minipage}\hfill
\begin{minipage}[b]{0.33676\linewidth}\centering \includegraphics[width=\linewidth,height=0.70\textheight,keepaspectratio]{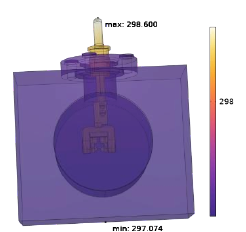}\par (b)\end{minipage}\hfill
\begin{minipage}[b]{0.31629\linewidth}\centering \includegraphics[width=\linewidth,height=0.70\textheight,keepaspectratio]{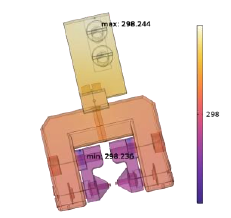}\par (c)\end{minipage}
\end{minipage}
\caption{~(a) Mesh of the FE model of the trap and vacuum chamber. (b) Simulated steady-state temperature distribution of the vacuum chamber and trap at 20.408 MHz and 500 V. (c) Simulated steady-state temperature distribution in the trap region.}\label{fig6}
\end{figure*}

Table \ref{tab3} summarizes the RF heating power dissipated in each trap component under this operating condition. The feedthrough ceramic dominates the total heat generation, accounting for approximately 94\% of the total RF loss, consistent with the heat-balance characteristics of low-heating end-cap trap designs \cite{m17}. The global energy balance of the FE model was confirmed by equating the total heat generation to the heat dissipation via external convection and radiation, yielding a residual of under 9\% that stems from numerical discretization. The calculated temperature distributions are illustrated in Fig. \ref{fig6}(b) and (c), indicating an ion-trap temperature of approximately 298.24 K.

\begin{table}[!htbp]
\centering
\caption{~RF heating power dissipated in each trap component under the nominal operating condition.}\label{tab3}
{\small\setlength{\tabcolsep}{3pt}
\begin{adjustbox}{max width=\linewidth}
\begin{tabular}{@{}lr@{}}
\toprule
Part & Power (mW) \\
\midrule
Feedthrough ceramic & 13.8 \\
Feedthrough kovar & 0.569 \\
Feedthrough copper rod & 0.28 \\
C-shaped copper carrier & $1.54\times10^{-3}$ \\
Alumina bolts & $1.36\times10^{-3}$ \\
Fused-silica spacer & $1.21\times10^{-3}$ \\
Cap electrode (Mo) & $2.49\times10^{-5}$ \\
RF electrode (Mo) & $2.93\times10^{-5}$ \\
Total heat generation & 14.55 \\
\bottomrule
\end{tabular}
\end{adjustbox}}
\end{table}
To evaluate the influence of mesh size, a mesh sensitivity analysis was performed for the key trap components at maximum element sizes of 2 mm, 1.5 mm and 1 mm, as summarized in Table \ref{tab4}. The RF port voltage amplitude remained stable across the three mesh levels with a variation below 0.001\%, confirming the reliability of the electromagnetic excitation. The dielectric losses in the feedthrough ceramic and fused-silica spacers varied by less than 0.3\%, indicating that the heat-source calculation is mesh-independent. The ultra-fine mesh with a maximum element size of 1 mm was therefore adopted as the reference discretization for the subsequent FE analyses, including the final evaluation of \textit{T}\textsubscript{eff}. The corresponding \ensuremath{\Delta}\textit{T}\textsubscript{eff} in the mesh-convergence test was 0.623 K. The difference between the two finest meshes, approximately 30 mK, was conservatively assigned as the mesh-discretization standard uncertainty, \textit{u}\textsubscript{mesh} = 31.0 mK.

\begin{table}[!htbp]
\centering
\caption{~Mesh sensitivity analysis of the FE model for maximum element sizes of 2, 1.5 and 1 mm, giving a mesh-induced uncertainty of 31.0 mK in \ensuremath{\Delta}\textit{T}\textsubscript{eff}.}\label{tab4}
{\small\setlength{\tabcolsep}{3pt}
\begin{adjustbox}{max width=\linewidth}
\begin{tabular}{@{}ccccc@{}}
\toprule
\shortstack{Max mesh size\\(mm)} & $V_{\mathrm{RF}}$ (V) & $Q_{\mathrm{ft}}$ (mW) & $Q_{\mathrm{spacer}}$ (mW) & $\Delta T_{\mathrm{ion}}$ (K) \\
\midrule
2 & 500.808 & 13.82 & $2.48\times10^{-3}$ & 0.533 \\
1.5 & 500.812 & 13.78 & $2.49\times10^{-3}$ & 0.593 \\
1 & 500.817 & 13.75 & $2.49\times10^{-3}$ & 0.623 \\
\bottomrule
\end{tabular}
\end{adjustbox}}
\end{table}
The geometric model represents the nominal trap configuration, whereas the actual trap is subject to machining and assembly tolerances. To evaluate their influence, the separation between the left and right end-cap electrodes and that between the left and right Mo RF electrodes were each varied by \ensuremath{\pm}0.05 mm about their nominal values, and the relative angular misalignment between the two Mo RF electrodes was varied by \ensuremath{\pm}5\ensuremath{{}^{\circ}}, while all other parameters were kept at their nominal values. The angular misalignment changed \ensuremath{\Delta}\textit{T}\textsubscript{eff} by up to 7.6 mK, compared with less than 3.2 mK for the spacing variations, indicating a greater sensitivity to electrode orientation than to spacing. Across all seven configurations, \ensuremath{\Delta}\textit{T}\textsubscript{eff} spanned 10.3 mK with a sample standard deviation of 3.7 mK. The individual configurations are listed in Supplementary Section \hyperref[supplement-sec-S2.5]{S2.5}, and the corresponding uncertainty contribution is included in Section \ref{main-sec-6.5}.

\begin{figure}[!htbp]
\centering
\hypertarget{main-fig-7}{}
\includegraphics[width=85.0000mm,keepaspectratio]{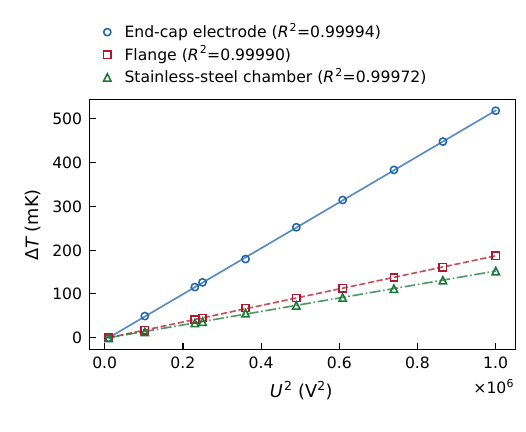}
\caption{~Simulated temperature rise \ensuremath{\Delta}T at three representative locations versus the square of the RF drive voltage amplitude U\textsuperscript{2}, with the external heat source at the copper rod disabled.}\label{fig7}
\end{figure}

Both ohmic and dielectric losses are approximately proportional to the square of the RF drive amplitude, so the RF-induced temperature rise is expected to follow an approximate U\textsuperscript{2} dependence. To verify this scaling, FE simulations were performed at ten RF voltage amplitudes (100, 320, 480, 500, 600, 700, 780, 860, 930, and 1000 V), with the fixed-temperature boundary at the top of the copper rod disabled to isolate the intrinsic RF-driven heating from the contribution of the helical resonator. The temperature rise \ensuremath{\Delta}T was extracted at three representative locations—the end-cap electrode, flange, and stainless-steel chamber, which span the main heat-conduction path from the electrode region to the chamber. As shown in Fig. \ref{fig7}, \ensuremath{\Delta}T at all three locations exhibits a strong linear dependence on U\textsuperscript{2}, with R\textsuperscript{2} values ranging from 0.99972 to 0.99994, confirming the expected quadratic scaling. Minor deviations from this ideal relationship stem from the fourth-power temperature dependence of thermal radiation \cite{m17}. 

\subsection{Parameter space and simulation dataset}\label{main-sec-3.3}
The FE model of Sections \ref{main-sec-3.1} and \ref{main-sec-3.2} is determined once its material and boundary inputs are assigned. These inputs define the parameter space over which the surrogate must be valid.

The thermal conductivities and relative permittivities were treated as fixed bulk properties with \ensuremath{\pm}10\% uncertainty ranges, contributing 9.83 mK to u(T\textsubscript{eff}); 97\% of this arises from the metallic thermal conductivities, while each permittivity contributes less than 0.5 mK (Section \ref{main-sec-6.5} and \hyperref[supplement-sec-S2.3]{S2.3}). In contrast, surface emissivity depends strongly on surface morphology, the oxidation state and temperature, making published values difficult to transfer directly to the assembled trap. Dielectric-loss tangents are even less transferable because they depend on the RF operating frequency and can vary appreciably between production batches of nominally identical insulating materials. This is particularly important here because dielectric loss in the insulating components is a major source of RF heating. They were therefore included in the sampled thermophysical parameter space and identified in Chapter \ref{main-sec-4}, while the remaining parameters were fixed at their reference values, with their uncertainties propagated as Type B contributions.

The surrogate uses eight physical inputs, but only five thermophysical quantities were treated as unknowns in the inverse problem. These comprise the emissivities of stainless steel, copper, and gold and the dielectric-loss tangents of the alumina feedthrough and fused-silica spacers. The FE training database additionally included the molybdenum emissivity. Although the Mo electrodes are gold coated on the ion-facing surfaces, a small number of non-ion-facing surfaces retain a molybdenum-emissivity boundary condition in the FE model. Variations in $\varepsilon_{\mathrm{Mo}}$ have only a weak influence on the predicted temperature field: a 100\% relative change in $\varepsilon_{\mathrm{Mo}}$ changes T\textsubscript{eff} by only 2.6 mK, compared with 97.9 mK for $\varepsilon_{\mathrm{SS}}$ (Supplementary Section \hyperref[supplement-sec-S2]{S2}). Therefore, $\varepsilon_{\mathrm{Mo}}$was retained in the surrogate training space to preserve its sensitivity direction in the reduced-basis construction of Section \ref{main-sec-3.4}, but was fixed during experimental inversion.

The remaining two inputs, the RF amplitude $V_{\mathrm{RF}}$ and ambient temperature $T_{\mathrm{env}}$, were treated as operating-condition variables rather than inverse parameters. They were varied during dataset generation so that a single surrogate could represent different RF drive and laboratory-temperature conditions, while their experimentally determined values were supplied directly during parameter identification and cross-condition validation. 

The corresponding reference values and sampling intervals are summarized in Table \ref{tab5}. For the dielectric-loss tangents, sampling was performed in logarithmic space because their possible values span more than one order of magnitude, whereas the emissivities, RF voltage, and ambient temperature were sampled in linear space.

Parameter combinations were generated by Latin hypercube sampling (LHS) \cite{m48} over the eight-dimensional space, and an automated parameter-sweep workflow was used to evaluate the corresponding FE cases and export the steady-state temperature-rise fields as surrogate-model targets. The training-set size was determined by progressively increasing the number of FE samples and monitoring the validation error until further samples produced no appreciable improvement. This resulted in 5400 training cases. An additional 675 cases were reserved for validation and 675 were retained as an independent test set, giving 6750 FE solutions in total. A fixed random seed of 42 was used to ensure reproducible data splitting. The validation set was used to determine the training-set size and select the final checkpoint, whereas the test set was evaluated only after the surrogate had been fixed.

\begin{table}[!htbp]
\centering
\caption{Parameter ranges used for Latin hypercube sampling of the FE model dataset. }\label{tab5}
{\small\setlength{\tabcolsep}{3pt}
\begin{adjustbox}{max width=\linewidth}
\begin{tabular}{@{}lrrr@{}}
\toprule
parameter & reference & min & max \\
\midrule
$\varepsilon_{\mathrm{SS}}$ & 0.25 & 0.2 & 0.3 \\
$\varepsilon_{\mathrm{Cu}}$ & 0.10 & 0.08 & 0.12 \\
$\varepsilon_{\mathrm{Au}}$ & 0.05 & 0.04 & 0.06 \\
$\varepsilon_{\mathrm{Mo}}$ & 0.162 & 0.13 & 0.194 \\
$\tan\delta_{\mathrm{Alumina}}$ & $1\times10^{-4}$ & 0.00001 & 0.0002 \\
$\tan\delta_{\mathrm{FS}}$ & $1.5\times10^{-5}$ & 0.000001 & 0.00003 \\
$V_{\mathrm{RF}}$ & 500 & 470 & 536 \\
$T_{\mathrm{env}}$ & 296.69 & 296.23 & 297.40 \\
\bottomrule
\end{tabular}
\end{adjustbox}}
\end{table}
\subsection{Surrogate-model construction and training}\label{main-sec-3.4}
A direct FE calculation of the ion-trap temperature field requires approximately 10 min, whereas the trained surrogate evaluates the same case within milliseconds. Repeated FE calculations are therefore impractical for iterative parameter identification. To reduce this computational cost while retaining the spatial temperature information required for inversion, a sensitivity-enriched reduced-order surrogate was developed. The model combines a multi-mode proper orthogonal decomposition (POD) representation of the FE temperature field with a quadratic regression baseline and a residual multilayer perceptron (MLP). A quadratic model captures the smooth leading-order parameter dependence, while the residual MLP accounts for the remaining coupled nonlinearities; its sigmoid linear unit (SiLU) activations keep the mapping continuously differentiable, which is required for stable sensitivity evaluation during inversion.

The surrogate architecture was determined by the dimensionality, regularity, and sensitivity requirements of the inverse identification problem. The input space contains only eight physical parameters, whereas each temperature field contains more than 4\ensuremath{\times}10\textsuperscript{5} valid pixels. Because the geometry, mesh, and boundary-condition layout are fixed, all FE solutions share the same spatial support and can therefore be represented using a common low-dimensional basis. Moreover, the RF-induced temperature rise is only approximately 0.5 K on a 300 K background, over which the T\textsuperscript{4} radiative term remains close to linear, allowing the leading parameter dependence to be captured by a low-order analytic model. The generation of FE training samples is computationally expensive, making a compact and sample-efficient model preferable to simply increasing network capacity. More importantly, inverse identification requires accurate parameter sensitivities as well as accurate temperature-field predictions. Weak sensitivity directions may otherwise be lost through basis truncation, inaccurate regression derivatives, or a training objective dominated by the overall field error. The surrogate was therefore constructed using a sensitivity-enriched fixed spatial basis together with a low-order regression baseline and a smooth residual network, so as to retain these weak parameter-dependent responses while keeping the model compact and continuously differentiable. These considerations motivate the sensitivity-enriched spatial reduction and the quadratic-residual parameter mapping described below.

The surrogate takes the eight-dimensional physical parameter vector

\begin{equation}
\mathbf{x}=\left[\varepsilon_{\mathrm{SS}},\varepsilon_{\mathrm{Cu}},\varepsilon_{\mathrm{Au}},\varepsilon_{\mathrm{Mo}},\tan\delta_{\mathrm{Al}_2\mathrm{O}_3},\tan\delta_{\mathrm{FS}},V_{\mathrm{RF}},T_{\mathrm{env}}\right]^{\mathrm{T}}
\label{main-eq-37}
\end{equation}
as input and predicts the RF-induced temperature-rise field \ensuremath{\Delta}T(r). Instead of directly regressing the temperature at every FE pixel, the high-dimensional field is first projected onto a fixed low-dimensional spatial basis. The surrogate therefore predicts only the mean temperature rise and a small number of modal coefficients, from which the complete field is reconstructed. Using a fixed linear basis also allows the field sensitivities to be reconstructed directly from the sensitivities of the predicted coefficients, without introducing an additional nonlinear decoder.

\begin{figure*}[!htbp]
\centering
\hypertarget{main-fig-8}{}
\includegraphics[width=175.0000mm,height=0.70\textheight,keepaspectratio]{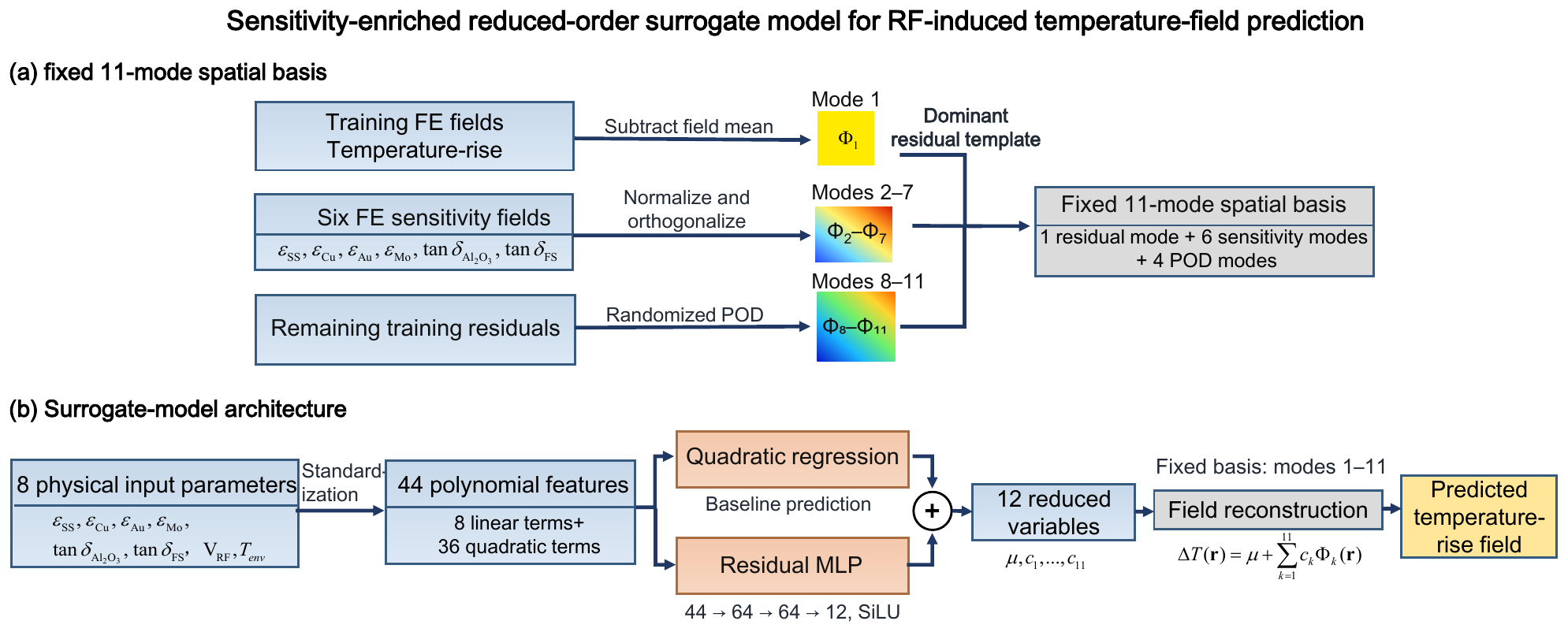}
\caption{~Architecture of the sensitivity-enriched reduced-order surrogate. (a) Construction of the 11-mode spatial basis from one dominant training-field mode, six orthogonal modes spanning the normalized FE-sensitivity subspace, and four additional POD modes extracted from the remaining training residuals. (b) Parameter-to-coefficient regression and reconstruction of the full temperature-rise field using the fixed spatial basis.}\label{fig8}
\end{figure*}

\textit{Spatial reduction}\textit{:}For each FE sample, the temperature rise over the fixed COMSOL valid-region mask is decomposed into its spatial mean \ensuremath{\mu} and a centered residual field. The latter is represented by an orthonormal basis according to

\begin{equation}
\Delta T(\mathbf{r})=\mu+\frac{1}{1000}\sum_{k=1}^{K}c_k\Phi_k(\mathbf{r}),
\label{main-eq-38}
\end{equation}
where $\Phi_k(\mathbf{r})$denotes the k-th spatial basis function and c\textsubscript{k} is its coefficient in mK. In the final model, K=11, so that an FE field containing more than 4\ensuremath{\times}10\textsuperscript{5} valid pixels is represented by only 12 reduced variables,

\begin{equation}
\mathbf{y}=[\mu,c_1,c_2,\ldots,c_{11}]^{\mathrm{T}}.
\label{main-eq-40}
\end{equation}
The reduced basis was designed not only to reproduce the dominant temperature distribution but also to preserve weak spatial directions that are important for inverse parameter identification \cite{m45}. As illustrated in Fig. \ref{fig8}(a), the basis consists of three components. The first mode is the dominant centered residual-temperature template obtained from the training fields. The next six modes span the FE-sensitivity subspace associated with the four surface emissivities and the two dielectric-loss tangents. Because the magnitudes of these six sensitivity fields differ by several orders of magnitude, each sensitivity field is normalized before singular-value decomposition and orthogonalization. This equalization prevents the strongest sensitivity direction from dominating the basis and preserves the weak parameter-dependent field structures. After projection onto these seven modes, four additional randomized-POD modes are extracted from the remaining training residuals \cite{m46}. The final basis therefore contains one dominant training-field mode, six sensitivity-enriched modes, and four residual POD modes.

\textit{Parameter-to-coefficient regression}\textit{.}~As shown in Fig. \ref{fig8}(b), the eight physical inputs are standardized and expanded into 44 first- and second-order polynomial features, comprising eight linear and 36 quadratic terms. A quadratic regression provides the baseline prediction of the twelve reduced variables, while a residual MLP with layer widths 44-64-64-12 learns the remaining nonlinear correction. This decomposition allows the smooth leading-order parameter dependence to be represented by the quadratic model, leaving only the residual nonlinear response to the MLP. SiLU activation functions \cite{m47} are used to maintain a smooth nonlinear mapping and facilitate stable evaluation of parameter sensitivities during inversion. The predicted mean temperature rise $\mu$and 11 modal coefficients $c_1,\ldots,c_{11}$are finally combined with the fixed spatial basis to reconstruct the full FE temperature field.

\textit{Training objective.}\textit{~}The spatial basis is orthonormal under the discrete Euclidean inner product, $\sum_{i=1}^{N_{\mathrm{pix}}}\phi_j(i)\phi_k(i)=\delta_{jk}$.Owing to the centered and orthonormal spatial basis, the full-field mean-square error can be expressed in reduced coordinates as

\begin{equation}
\mathcal{L}_{\mathrm{field}}=(1000\Delta\mu)^2+\frac{1}{N_{\mathrm{pix}}}\sum_{k=1}^{11}(\Delta c_k)^2,
\label{main-eq-45}
\end{equation}
where N\textsubscript{pix} is the number of valid FE pixels. The sensitivity term penalizes the normalized difference between the surrogate and FE sensitivity fields,

\begin{equation}
\mathcal{L}_{sens}=\frac{1}{6}\sum_{j=1}^{6}\left\lVert\frac{\partial\Delta\hat{T}/\partial q_j-\partial\Delta T_{FE}/\partial q_j}{s_j}\right\rVert_2^2,
\label{main-eq-46}
\end{equation}
where \textit{q}\textsubscript{j} denotes the normalized value of the j-th thermophysical parameter and \textit{s}\textsubscript{j}

is the characteristic magnitude of its FE sensitivity field. The total training objective is

\begin{equation}
\mathcal{L}=\mathcal{L}_{\mathrm{field}}+\lambda_{\mathrm{sens}}\mathcal{L}_{\mathrm{sens}}.
\label{main-eq-47}
\end{equation}
where $\lambda_{\mathrm{sens}}$ is the weighting coefficient for the sensitivity loss. The parameter-wise normalization in Eq.~\eqref{main-eq-46} prevents the dominant sensitivity direction from overwhelming weaker responses, thereby preserving the information required for multi-parameter inversion. 

The residual MLP was trained with the AdamW optimizer \cite{m49}; the hyperparameters, learning-rate schedule, early-stopping criterion, and the setting of the sensitivity-loss weight $\lambda_{\mathrm{sens}}$ are listed as Table \hyperref[tabS1]{S1} in the Supplementary Material. To evaluate the influence of neural-network initialization, three independent initializations were evaluated, and the final checkpoint selection was based on the combined validation criterion incorporating both reconstructed-field accuracy and FE-sensitivity fidelity. 

\subsection{Prediction accuracy and sensitivity fidelity}\label{main-sec-3.5}
The three independent initializations gave comparable validation performance. Seed 2024 gave the lowest combined validation criterion and was therefore selected at its epoch-290 checkpoint. The individual initialization results are given in Supplementary Section \hyperref[supplement-sec-S1.1]{S1.1}.  

Additional architecture comparisons and ablation tests are provided in Supplementary Section \hyperref[supplement-sec-S1.1]{S1.1}. These results confirm that the reduced-order POD representation is important for preserving parameter sensitivities, while the quadratic baseline and residual MLP further improve both field-prediction accuracy and sensitivity fidelity.

\begin{figure*}[!htbp]
\centering
\hypertarget{main-fig-9}{}
\includegraphics[width=170.0000mm,keepaspectratio]{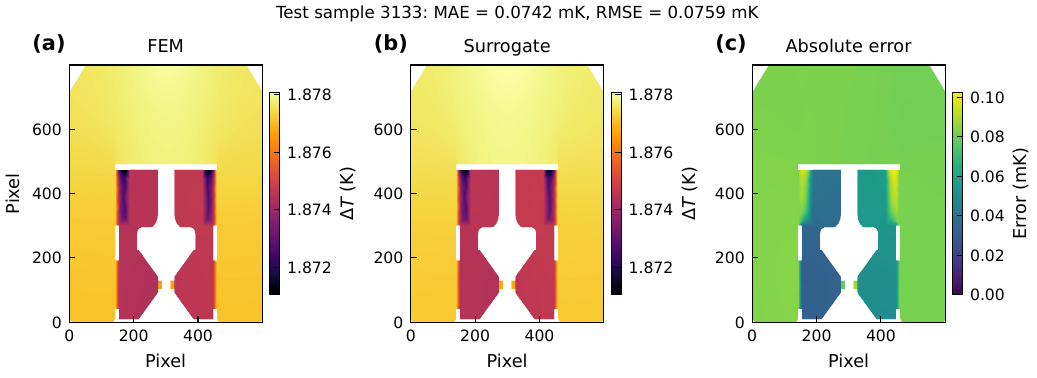}
\caption{Comparison between the FE and surrogate temperature-rise fields for a test case close to the 95th-percentile prediction error.}\label{fig9}
\end{figure*}

Fig. \ref{fig9} shows a near-worst case, selected at the 95th percentile of the test MAE distribution. Its full-field MAE and RMSE are 0.07422 and 0.07594 mK, and the maximum absolute error in the valid region is 0.1022 mK. Even in this case, the surrogate reproduces both the overall temperature rise and the localized spatial gradients of the FE solution. Over the whole test set the correlation coefficient between the centered FE and surrogate fields averages 0.999994. Because the spatial mean is removed before the correlation is formed, this measures the preservation of the non-uniform structure rather than of the overall level. 

Across all valid pixels in the 675 independent test fields, the MAE and RMSE were 0.02220 mK and 0.03511 mK, with a median per-field MAE of 0.01250 mK and 95\% of fields below 0.07432 mK. The mean field bias over the test set was -2.042\ensuremath{\times}10\textsuperscript{-4} mK, indicating no significant systematic over- or underestimation. The complete error distribution and mean-field bias are shown in Supplementary Fig. \hyperlink{supplement-fig-S1}{S1}. 

To verify that the eleven-modes basis was sufficient to represent the FE temperature fields, each test field was directly projected onto the fixed basis and reconstructed. The resulting MAE and RMSE were 5.507\ensuremath{\times}10\textsuperscript{-5} mK and 1.882\ensuremath{\times}10\textsuperscript{-4} mK, roughly two orders of magnitude below the surrogate's prediction error, so the reduced representation is not the limiting factor. The residual error is attributable to the parameter-to-coefficient regression rather than to the truncation. 

\begin{figure*}[!htbp]
\centering
\hypertarget{main-fig-10}{}
\includegraphics[width=170.0000mm,keepaspectratio]{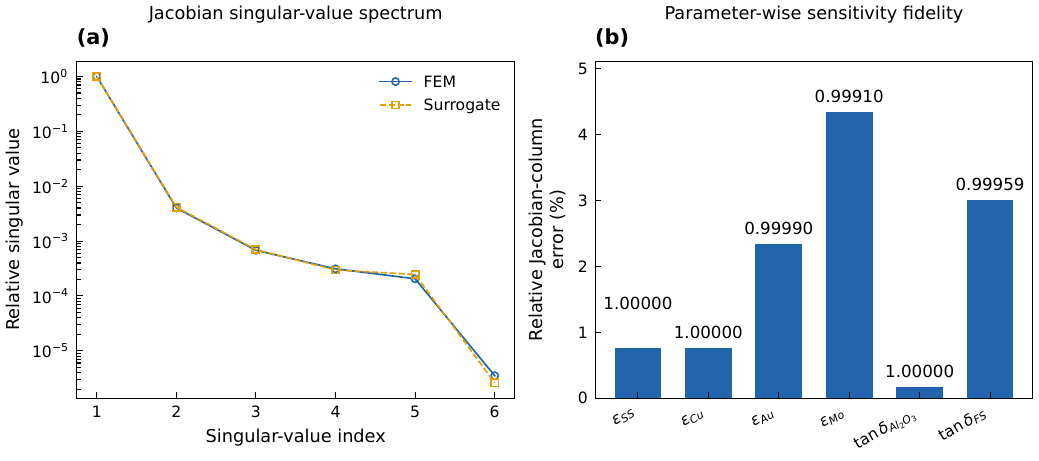}
\caption{~Fidelity of the surrogate-model parameter sensitivities with respect to the FE model. (a) Relative singular-value spectra of the normalized six-parameter Jacobians obtained from the FE model and the surrogate. (b) Relative errors between the FE and surrogate sensitivity fields for each parameter. The numbers above the bars denote the cosine similarity between the corresponding FE and surrogate sensitivity fields.}\label{fig10}
\end{figure*}

Because the surrogate is ultimately used for inverse identification, accurate field prediction alone is not sufficient. The local parameter sensitivities of the trained surrogate were therefore compared with those of the FE model at the reference parameter set. Fig. \ref{fig10}(a) compares the singular-value spectra of the normalized six-parameter sensitivity matrices (Jacobians). The FE and surrogate spectra overlap closely over more than five orders of magnitude, and the relative Frobenius error between the complete FE and surrogate sensitivity matrices is 0.2225\%. Fig. \ref{fig10}(b) compares the sensitivity associated with each parameter individually: the relative errors range from 0.1605\% to 4.333\%, while the cosine similarities all exceed 0.9991. The surrogate therefore preserves both the magnitude and the spatial pattern of the parameter sensitivities required for inverse identification. 

The agreement in both temperature fields and parameter sensitivities is essential for reliable inverse identification, because the six thermophysical parameters affect the temperature field with very different strengths. By preserving the weak sensitivity directions while keeping the prediction error well below 1 mK, the surrogate can be used for repeated radiance-domain inversion without rerunning the FE model at each iteration. The residual 4.333\% worst-case sensitivity discrepancy is carried into the budget in Section \ref{main-sec-6.5}.

\section{Inverse identification method}\label{main-sec-4}
The surrogate developed in Section \ref{main-sec-3} predicts the surface temperature-rise field for a given set of thermophysical parameters, whereas the inverse problem determines the parameter values that reproduce the measured radiance distribution. After selection of the final surrogate checkpoint, the final surrogate model was frozen. The experimental analysis was then divided into two stages: the broadband dataset acquired at 500 V was used for thermophysical-parameter identification using the component-wise registered radiance field and predefined interior quantitative supports, while a second broadband dataset acquired at the lower RF voltage of 485 V was used for independent predictive validation without parameter refitting.

\begin{figure}[!htbp]
\centering
\hypertarget{main-fig-11}{}
\includegraphics[width=85.0000mm,height=0.70\textheight,keepaspectratio]{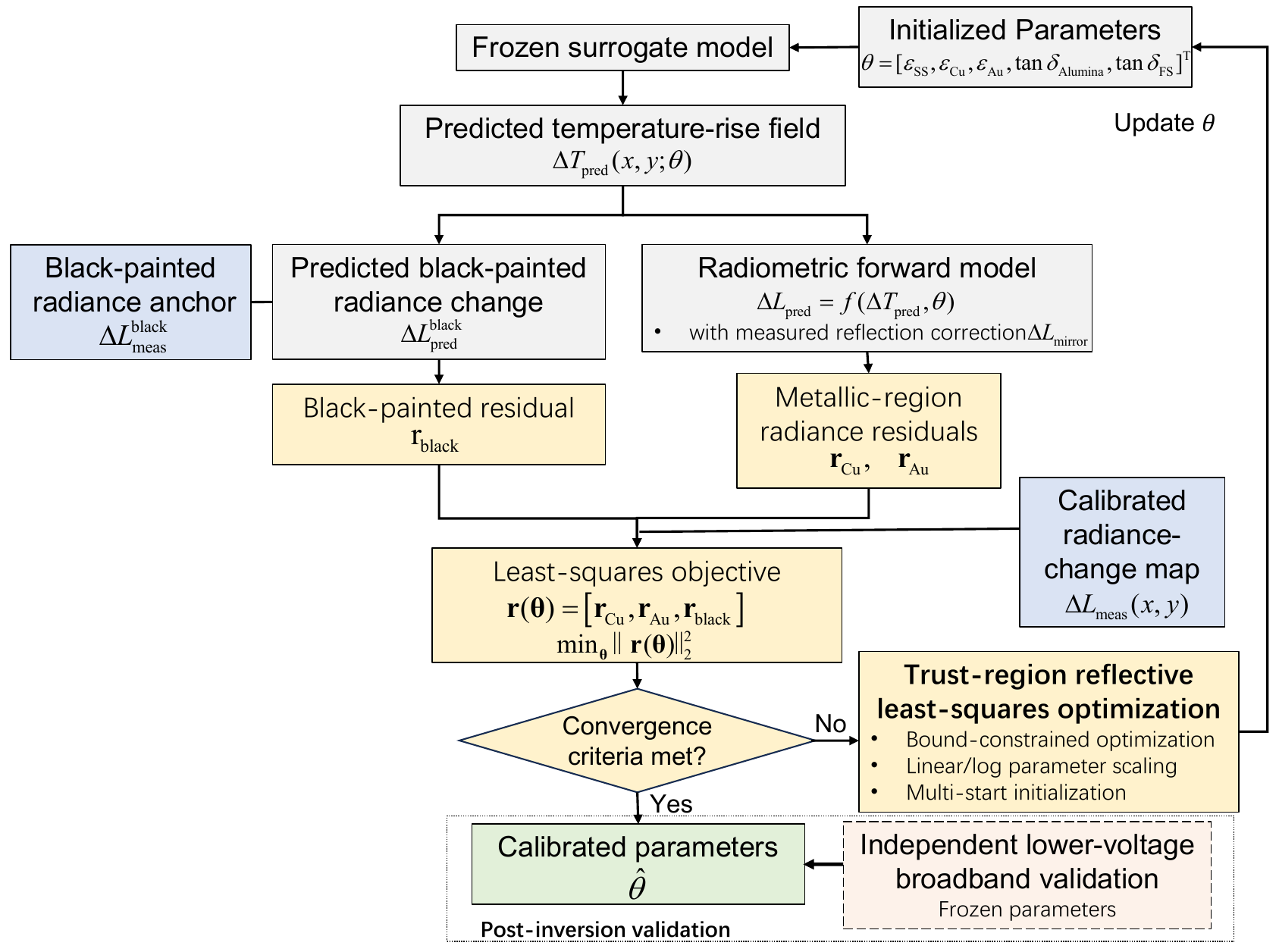}
\caption{~Workflow of radiance-domain parameter identification using the frozen surrogate. For a candidate parameter, the surrogate-predicted RF temperature-rise field is converted into a radiance-change distribution using the radiometric forward model with measured reflection correction. Residuals from the Cu, Au, and black-painted regions are jointly minimized by bounded trust-region reflective least-squares optimization to identify the thermophysical parameters. The retrieved parameters are then frozen for independent lower-voltage broadband validation without refitting.}\label{fig11}
\end{figure}
\subsection{Identified parameters and parameter bounds}\label{main-sec-4.1}
Of the eight surrogate inputs, five thermophysical parameters are treated as unknowns and identified from the infrared measurements:

\begin{equation}
\theta=\left[\varepsilon_{\mathrm{SS}},\varepsilon_{\mathrm{Cu}},\varepsilon_{\mathrm{Au}},\tan\delta_{\mathrm{Al}_2\mathrm{O}_3},\tan\delta_{\mathrm{FS}}\right]^{\mathrm{T}},
\label{main-eq-53}
\end{equation}
Which comprises the emissivities of stainless steel, copper, and gold, together with the dielectric-loss tangents of the alumina feedthrough and fused-silica spacers. The molybdenum emissivity is held at its reference value. The RF amplitude and ambient temperature were supplied as measured operating-condition inputs rather than adjusted.

The five parameters are transformed into normalized coordinates q bounded within [\ensuremath{-}1, 1] over the ranges of Table \ref{tab5}, linearly for the emissivities and logarithmically for the two dielectric-loss tangents, matching the sampling of Section \ref{main-sec-3.3}. This scaling places the five inverse parameters on a comparable numerical scale and restricts the optimization to the parameter ranges represented in the surrogate training set.

\subsection{Radiance-domain objective function and weighting}\label{main-sec-4.2}
For each candidate parameter vector $\theta$, the frozen surrogate generates the RF-induced temperature-rise field.

\begin{equation}
\Delta T_{\mathrm{RF}}\left(\mathbf{r};\theta,V_{\mathrm{RF}},T_{\mathrm{env}}\right).
\label{main-eq-55}
\end{equation}
The radiometric forward model then converts the field into the radiance-change distribution that the camera would record. The emission contribution predicted for material region $m$ is calculated as

\begin{equation}
\begin{aligned}
\Delta L_m^{\mathrm{pred}}(\mathbf{r})&=\varepsilon_m\bigl\{B_{7\text{--}14}\left[T_{\mathrm{env,on}}+\Delta T_{\mathrm{RF}}(\mathbf{r})\right]\\
&\quad -B_{7\text{--}14}\left(T_{\mathrm{env,off}}\right)\bigr\}\\
&\quad +(1-\varepsilon_m)\Delta L_{\mathrm{refl},m},
\end{aligned}
\label{main-eq-57}
\end{equation}
where $B_{7\text{--}14}(T)$ is the Planck radiance integrated over the nominal 7–14 \ensuremath{\mu}m camera band and $\Delta L_{\mathrm{refl},m}$ denotes the component-dependent reflected-radiance contribution anchored by the silver-mirror measurement. As described in Section \ref{main-sec-2.3.3}, RF-on/off subtraction removes most of the static reflected background, but a residual reflected-radiance change remains when the surrounding radiative field differs between the two states. Infrared mirrors can be used to estimate reflected ambient radiation \cite{m50}. A silver mirror with a manufacturer-specified reflectance exceeding 96\% was therefore recorded through the same optical path under the RF-on and RF-off states. The mirror was fixed throughout the measurements, with no change in its position or orientation. The correction was represented separately over the Cu, Au-cap, Au-end, and black-painted supports because these surfaces differ in orientation, depth, and radiative view.

Temperature–emissivity inference from radiance is underdetermined without additional constraints \cite{m51}. Here, the black-painted region has a known, fixed emissivity, so its radiance change constrains the amplitude of the temperature-rise field directly and reduces that degeneracy. The Cu, Au-cap and Au-end regions carry the unknown emissivities and supply the sensitivity from which those emissivities and the loss tangents are identified. 

For each radiometric group $m$, the uncertainty-normalized pixel residual is defined as
\begin{equation}
\mathbf{r}_{m,i}=\frac{\Delta L_{m,i}^{\mathrm{pred}}-\Delta L_{m,i}^{\mathrm{exp}}}{\sigma_m\sqrt{N_m}}
\label{main-eq-61}
\end{equation}
,
where $\sigma_m$ is the standard uncertainty derived from the temporal repeatability of the measured radiance difference. For Cu and the black-painted region, $N_m$ denotes the corresponding number of valid pixels. For Au-cap and Au-end, both subregions were normalized using the total Au pixel count, $N_{\mathrm{Au}}=N_{\mathrm{Au,cap}}+N_{\mathrm{Au,end}}$, so that separating the Au surface into two radiometric groups did not artificially increase its overall weight in the inversion objective. The normalization by $N_m$ prevents a radiometric group from dominating the objective simply because it contains more valid pixels. The normalization by $\sigma_m$ expresses each residual relative to its own measurement noise. These noise scaling and group-balancing factors are choices of the present inversion objective.

Statistical uncertainties in radiometric temperature and emissivity estimation have been analysed using least-squares and covariance methods \cite{m52}. Here, least-squares optimization combined with the radiometric forward model is used to estimate the thermophysical parameters. The inverse objective is then written as

\begin{equation}
\mathcal{J}(\boldsymbol{\theta})=\sum_{m\in\{\mathrm{Cu},\mathrm{Au},\mathrm{black}\}}\lVert\mathbf{r}_m\rVert_2^2.
\label{main-eq-67}
\end{equation}
The resulting objective was minimized using bounded trust-region reflective least-squares optimization \cite{m53,m54}. A single descent establishes a local minimum only, and the conditioning of the problem is not known in advance, so the minimization was repeated from twenty-four independent initializations distributed within the prescribed bounds \cite{m55}.

\subsection{Identification setup and solution selection}\label{main-sec-4.3}
For the identification condition, the ion trap was operated at an RF frequency of 20.408 MHz with a voltage amplitude of 500 V, consistent with the FE model. Seven RF-on thermograms and seven RF-off thermograms were averaged separately. The measured ambient temperatures were 297.283 K under the RF-on condition and 297.217 K under the RF-off condition. The averaged raw images were converted into broadband radiance using the calibration relation in Section \ref{main-sec-2.3.2}. The RF-on and RF-off images were then spatially registered using the procedure described in Section \ref{main-sec-2.3.3} and subtracted in the linear radiance domain to obtain the measured radiance-change field.

A fixed central circular ROI of radius 0.8 R, selected within the stable range of the ROI-size robustness scan, was used to characterize the mirror response. For the identification condition, the mean radiance change within this ROI was

\begin{equation}
\Delta L_{\mathrm{mirror}}=0.1768\,\mathrm{W}\,\mathrm{m}^{-2}\mathrm{sr}^{-1}.
\label{main-eq-68}
\end{equation}
For this identification condition, the 500-V RF operating condition and the mean ambient temperature of 297.250 K were supplied to the surrogate as known operating-condition inputs.

To account for parallax and perspective differences among the visible trap surfaces, the measured radiance-change field was mapped to the native surrogate grid using six component-specific four-corner projective transformations. Separate transformations were used for the camera-left and camera-right parts of the Cu, Au and black-painted regions. For each component, the corresponding projective transformation was applied consistently to both the measured radiance-change field and its camera-space material mask.

The quantitative support for each component was defined by the intersection of the transformed experimental mask and the corresponding native surrogate component mask. To suppress mixed pixels and boundary-related registration errors, only pixels lying at least 8 native-surrogate pixels inside both boundaries were retained. Pixels near visible Cu ridges and edges that were not represented in the idealized surrogate geometry were additionally excluded using masks defined from the RF-off thermogram to avoid bias from local geometric features present in the experiment but absent from the FE/surrogate model. Pixels assigned to more than one transformed component were also excluded. After applying the component-wise registration and interior-support criteria, the quantitative supports contained 110166 Cu pixels, 6587 Au-cap pixels, 50099 Au-end pixels, and 46494 black-painted pixels.

Where the starts do not agree, the nominal solution is selected by three criteria applied in order: the lowest objective value; a position in the interior of the bounds rather than against one, since a solution resting on a bound reflects the sampling range as much as the data; and the largest basin of attraction among the starts. Alternative solutions not adopted as nominal are retained rather than discarded when their residuals cannot distinguish them from the selected solution; the resulting separation in \textit{T}\textsubscript{eff} is carried into the uncertainty budget in Section \ref{main-sec-6.3}.

\subsection{Identifiability and convergence diagnostics}\label{main-sec-4.4}
To assess the robustness of the inverse identification, 24 independent initializations were tested within the prescribed parameter ranges. Twenty-three of the 24 runs converged to a tightly clustered parameter set, while one initialization yielded a distinct alternative solution. None of the 24 terminal solutions reached the near-bound criterion. The terminal weighted RMS values were nearly indistinguishable, with a minimum of 0.01605814 and a total span of only 9.025\ensuremath{\times}10\textsuperscript{-10} (Fig. \ref{fig12}(a)). The full multistart results are given in Supplementary Section \hyperref[supplement-sec-S2.1]{S2.1}.

The multistart results therefore indicate stable numerical convergence together with a weak residual parameter non-uniqueness. Propagating the 24 terminal parameter vectors through the surrogate gives a maximum pointwise standard deviation of only 0.009962 mK in the reconstructed temperature-rise field, showing that the alternative parameter combination produces an essentially indistinguishable temperature-field prediction. The most prominent variation occurs in the stainless-steel emissivity. This is physically because the stainless-steel enclosure is not directly observed by the camera, so its emissivity is constrained only indirectly through its influence on the radiative environment of the observed trap surfaces. In contrast, the Cu and Au emissivities remain nearly unchanged between the main solution cluster and the alternative solution.

The solution from start 23 was adopted as the nominal solution because it gave the minimum objective value and remained within the prescribed parameter bounds. The isolated alternative solution was retained as admissible because its objective value was nearly indistinguishable from that of the main solution cluster; its effect on \textit{T}\textsubscript{eff} is therefore propagated as a multistart non-uniqueness contribution in Section \ref{main-sec-6.3}.

\section{Parameter-identification results and independent validation}\label{main-sec-5}
\subsection{Inferred parameters and consistency checks}\label{main-sec-5.1}
The identified thermophysical parameters are given in Table \ref{tab6}. The final normalized coordinates were

\begin{equation}
q=[0.1031,-0.3208,-0.1226,0.5757,0.2515],
\label{main-eq-69}
\end{equation}
and none of the five parameters reached the near-bound criterion $|q|\geq0.95$.

corresponding to

\begin{equation}
\varepsilon_{\mathrm{SS}}=0.2552,\varepsilon_{\mathrm{Cu}}=0.09358,\varepsilon_{\mathrm{Au}}=0.04877,
\label{main-eq-71}
\end{equation}
\begin{equation}
\tan\delta_{\mathrm{Al}_2\mathrm{O}_3}=1.059\times10^{-4},\tan\delta_{\mathrm{FS}}=8.400\times10^{-6}.
\label{main-eq-72}
\end{equation}
The identified parameter set was then frozen and used without adjustment in the independent lower-voltage broadband validation.

The retrieved stainless-steel emissivity of 0.2552 is comparable in magnitude to 0.2746, calculated at 23 $^{\circ}$C from the linear fit for machined 304 stainless steel in Table 1 of Ablewski et al. \cite{m38}. Their measured total hemispherical emissivity and the band-limited radiometric constraint used here are different quantities, so this comparison is qualitative. The copper emissivity may reflect differences in surface condition and oxidation \cite{m36}. These comparisons do not independently validate the inversion.

\begin{table}[!htbp]
\centering
\caption{~Thermophysical parameters identified from the radiance-domain inversion.}\label{tab6}
{\small\setlength{\tabcolsep}{3pt}
\begin{adjustbox}{max width=\linewidth}
\begin{tabular}{@{}llr@{}}
\toprule
Parameter & Treatment & Final value \\
\midrule
$\varepsilon_{\mathrm{SS}}$ & identified & 0.2552 \\
$\varepsilon_{\mathrm{Cu}}$ & identified & 0.09358 \\
$\varepsilon_{\mathrm{Au}}$ & identified & 0.04877 \\
$\tan\delta_{\mathrm{Al}_2\mathrm{O}_3}$ & identified & 1.059\ensuremath{\times}10\textsuperscript{-}\textsuperscript{4} \\
$\tan\delta_{\mathrm{FS}}$ & identified & 8.400\ensuremath{\times}10\textsuperscript{-}\textsuperscript{6} \\
$\varepsilon_{\mathrm{black}}$ & Calibrated radiometric reference & 0.82 \\
\bottomrule
\end{tabular}
\end{adjustbox}}
\end{table}
\subsection{Reconstruction of broadband thermograms and spatial residuals}\label{main-sec-5.2}
\begin{figure*}[!htbp]
\centering
\hypertarget{main-fig-12}{}
\includegraphics[width=170.0000mm,height=0.70\textheight,keepaspectratio]{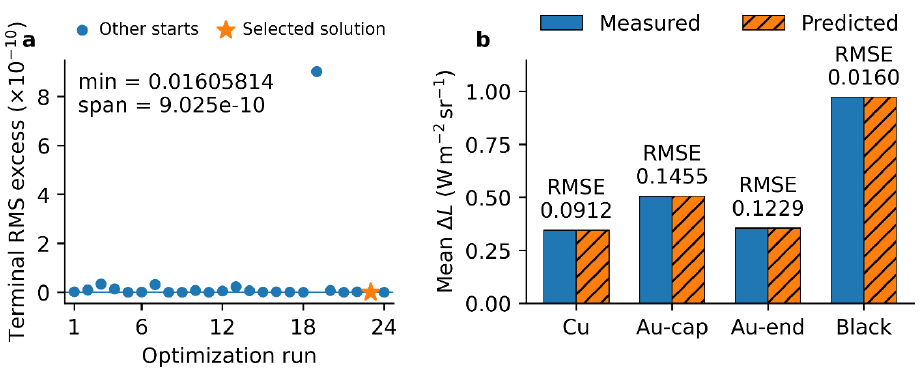}
\caption{~Multistart convergence and identification fit. (a) Terminal weighted-RMS excess relative to the minimum for twenty-four independent initializations; the star indicates the selected solution. (b) Measured and predicted mean radiance changes for the Cu, Au-cap, Au-end and black-painted regions under the identification condition; the values above the bars are the corresponding pixelwise RMSEs in W m\textsuperscript{-}\textsuperscript{2} sr\textsuperscript{-}\textsuperscript{1}.}\label{fig12}
\end{figure*}

Under the identification condition the measured and predicted regional mean radiance changes were 0.3451 and 0.3451 W m\textsuperscript{-2}sr\textsuperscript{-1} for Cu, 0.5036 and 0.5050 for Au-cap, 0.3554 and 0.3553 for Au-end, and 0.9712 and 0.9712 for the black-painted region. The corresponding mean radiance residuals between prediction and measurement regional mean radiance changes were -1.109\ensuremath{\times}10\textsuperscript{-6}, 1.394\ensuremath{\times}10\textsuperscript{-}\textsuperscript{3} , -1.399\ensuremath{\times}10\textsuperscript{-4} and -1.687\ensuremath{\times}10\textsuperscript{-6} W m\textsuperscript{-2}sr\textsuperscript{-1}, respectively. The corresponding pixelwise RMSEs were 0.09121, 0.1455, 0.1229 and 0.01601 W m\textsuperscript{-2} sr\textsuperscript{-1}. (Fig. \ref{fig12}(b)).

The regional mean radiance changes were reproduced to within 0.3\% for all four regions. The remaining pixelwise residuals were smallest over the black-painted region and largest over Au-cap, indicating that the principal discrepancies are spatially localized rather than associated with a common error in the predicted radiance-change amplitude. These residual structures are consistent with local reflected-radiance variations and geometric details that are not fully represented by the idealized FE/surrogate model.

\subsection{Cross-condition validation at a lower RF voltage}\label{main-sec-5.3}
The transferability of the identified parameter set was tested on a broadband dataset acquired at a lower RF voltage amplitude of 485 V. The RF-off sequence used for the identification condition was retained as the common RF-off reference. The measured ambient temperatures were 297.342 K under RF-on and 297.217 K under RF-off. Their mean value, 297.280 K, was supplied to the surrogate. The component-dependent reflected-radiance representation established for the identification condition was retained unchanged, while the mirror-derived reference amplitude was updated to 0.1555 W m\textsuperscript{-2} sr\textsuperscript{-1} for the validation condition.

All identified thermophysical parameters, the molybdenum and black-paint emissivities, the material masks, the component-wise projective transformations, the quantitative supports, and the radiometric forward-model formulation were retained unchanged. Only the RF amplitude, the measured ambient temperature and the silver-mirror radiance change corresponding to the validation condition were updated. No thermophysical parameter was refitted.

For each region $m$,the standardized mean residual is

\begin{equation}
z_m=\frac{\overline{\Delta L}_m^{\mathrm{pred}}-\overline{\Delta L}_m^{\mathrm{exp}}}{\sigma_m}.
\label{main-eq-81}
\end{equation}
and the overall cross-condition score was

\begin{equation}
z_{\mathrm{RMS}}=\sqrt{\frac{z_{\mathrm{Cu}}^2+z_{\mathrm{Au,cap}}^2+z_{\mathrm{Au,end}}^2+z_{\mathrm{black}}^2}{4}}.
\label{main-eq-82}
\end{equation}
Here, $z_m$ expresses the regional mean prediction error relative to the temporal-repeatability standard uncertainty, while $z_{\mathrm{RMS}}$ summarizes the agreement over the four radiometric regions.

\begin{figure*}[!htbp]
\centering
\hypertarget{main-fig-13}{}
\begin{minipage}{0.78\linewidth}\centering
\includegraphics[width=0.96000\linewidth,height=0.70\textheight,keepaspectratio]{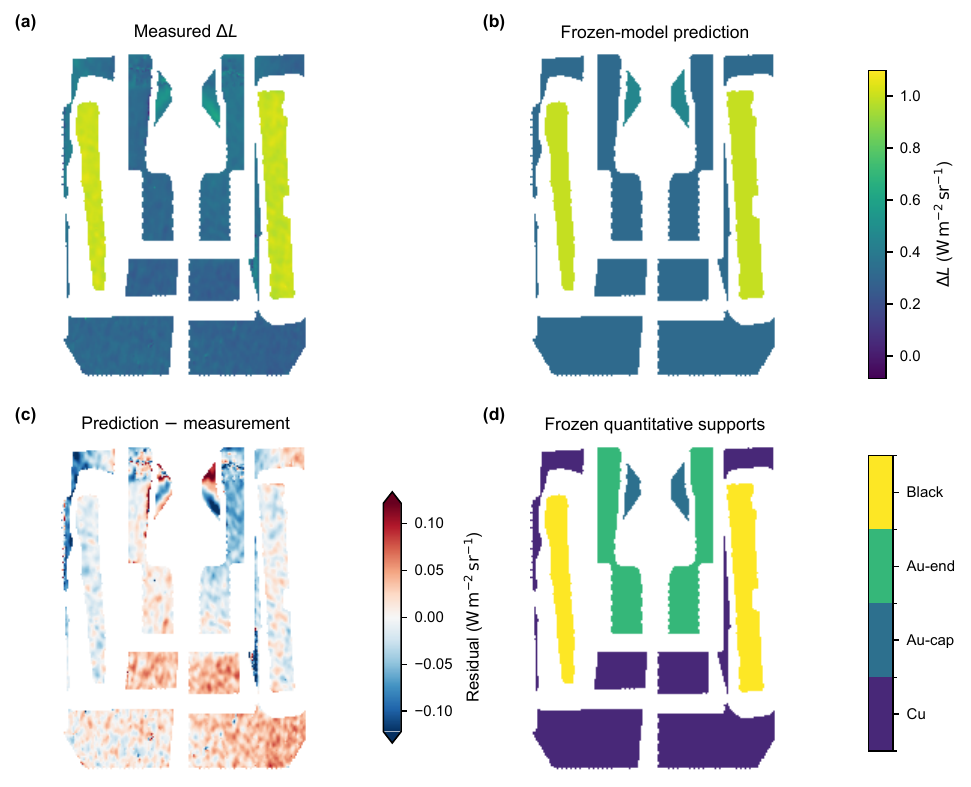}
\end{minipage}
\caption{~Cross-condition lower-voltage validation using the frozen identification result. (a) Measured radiance-change distribution. (b) Prediction obtained with the frozen thermophysical parameters and component transformations. (c) Prediction-minus-measurement residual. (d) Frozen quantitative supports corresponding to the Cu, Au-cap, Au-end, and black-painted regions.}\label{fig13}
\end{figure*}

The validation results are summarized in Table \ref{tab7} and Fig. \ref{fig13}. For Cu, the measured and predicted mean radiance changes were 0.3095 and 0.3219 W m\textsuperscript{-2}sr\textsuperscript{-1}, respectively, giving a mean error of +0.01244 W m\textsuperscript{-2}sr\textsuperscript{-1}, a pixelwise RMSE of 0.03912 W m\textsuperscript{-2}sr\textsuperscript{-1}, and a standardized residual of +0.544. For Au-cap, the corresponding measured and predicted values were 0.4608 and 0.4568 W m\textsuperscript{-2}sr\textsuperscript{-1}, giving a mean error of \ensuremath{-}0.003948 W m\textsuperscript{-2}sr\textsuperscript{-1}, a pixelwise RMSE of 0.07447 W m\textsuperscript{-2}sr\textsuperscript{-1}, and a standardized residual of \ensuremath{-}0.168. For Au-end, the measured and predicted values were 0.3379 and 0.3237 W m\textsuperscript{-2}sr\textsuperscript{-1}, giving a mean error of \ensuremath{-}0.01416 W m\textsuperscript{-2}sr\textsuperscript{-1}, a pixelwise RMSE of 0.03887 W m\textsuperscript{-2}sr\textsuperscript{-1}, and a standardized residual of \ensuremath{-}0.626. For the black-painted region, the measured and predicted values were 1.002 and 0.9959 W m\textsuperscript{-2}sr\textsuperscript{-1}, giving a mean error of \ensuremath{-}0.006265 W m\textsuperscript{-2}sr\textsuperscript{-1}, a pixelwise RMSE of 0.01737 W m\textsuperscript{-2}sr\textsuperscript{-1}, and a standardized residual of \ensuremath{-}0.279. The overall score was $z_{\mathrm{RMS}}=0.4455$. All four regional mean radiance changes were therefore reproduced within one temporal-repeatability standard uncertainty without parameter refitting.

The residuals have mixed signs, indicating no common bias across the four regional responses. As shown in Fig. \ref{fig13}(c), the largest spatial residual occurs over Au-cap, consistent with the local reflected-radiance and geometric effects observed under the identification condition. The remaining prediction error from this cross-condition validation is propagated as an uncertainty contribution in Section \ref{main-sec-6.3}.

\begin{table}[!htbp]
\centering
\caption{~Cross-condition validation at 485 V for the Cu, Au-cap, Au-end and black-painted regions, using the frozen identification result.}\label{tab7}
{\fontsize{8.5}{10.5}\selectfont\setlength{\tabcolsep}{3pt}
\begin{adjustbox}{max width=\linewidth}
\begin{tabular}{@{}lrrrrr@{}}
\toprule
Region & Measured & Predicted & Error & RMSE & $z_m$ \\
\midrule
Cu & 0.309467 & 0.321909 & +0.012442 & 0.039124 & +0.544 \\
Au-cap & 0.460789 & 0.456841 & \ensuremath{-}0.003948 & 0.074466 & \ensuremath{-}0.168 \\
Au-end & 0.337879 & 0.323720 & \ensuremath{-}0.014159 & 0.038865 & \ensuremath{-}0.626 \\
Black & 1.002192 & 0.995927 & \ensuremath{-}0.006265 & 0.017369 & \ensuremath{-}0.279 \\
Total & \multicolumn{5}{c}{$z_{\mathrm{RMS}}=0.4455$} \\
\bottomrule
\end{tabular}
\end{adjustbox}}
\end{table}
\subsection{Conduction-path validation}\label{main-sec-5.4}
The heat-conduction pathway predicted by the FE model was independently validated using contact thermometry on the external surface of the vacuum chamber. Four of the calibrated Pt100 sensors described in Section \ref{main-sec-2.2} were installed at four representative locations: adjacent to the ZnSe viewport flange, near the feedthrough flange, and on the side wall and bottom surfaces of the stainless-steel shell. These locations sample different parts of the heat-transfer network, including the region close to the dominant feedthrough heat source, the viewport-flange region, and chamber surfaces remote from the heat source. The temperature rise \ensuremath{\Delta}T at each location was recorded continuously for 1000 s and averaged to reduce random fluctuations.

The differences between the FE-predicted and measured temperature rises, defined as \ensuremath{\Delta}T\textsubscript{FE}\ensuremath{-}\ensuremath{\Delta}T\textsubscript{meas}, were \ensuremath{-}79.0, \ensuremath{-}7.1, +5.7, and \ensuremath{-}24.0 mK at the four locations, respectively. Three of the four locations agree within 25 mK, indicating that the FE model reproduces the overall magnitude and spatial distribution of heat conduction through the chamber. The largest discrepancy occurs at the ZnSe viewport flange. This region contains a mechanically assembled interface whose effective thermal conductance can depend on surface roughness, contact pressure, and assembly conditions, whereas the nominal FE model assumes ideal thermal contact at this interface. The larger local residual is therefore consistent with the greater modelling sensitivity expected at such an interface, although it cannot be uniquely attributed to contact resistance alone.

Overall, the comparison provides an experimental validation of the heat-transfer pathways represented by the FE model. However, the Pt100 sensors measure temperatures on the external chamber surface rather than the radiative environment at the ion position. Consequently, the maximum surface residual of 79.0 mK is not directly assigned as an uncertainty of T\textsubscript{eff}. Instead, the individual FE-related sources that can affect T\textsubscript{eff}, including mesh discretization, material properties, thermal-contact conductance, geometric tolerances, and thermal boundary conditions, are evaluated and propagated separately in the uncertainty budget of Section \ref{main-sec-6}.

\section{Effective BBR temperature and uncertainty evaluation}\label{main-sec-6}
\subsection{Effective BBR temperature at the ion position}\label{main-sec-6.1}
The identified parameters were reintroduced into the original coupled FE model, and \textit{T}\textsubscript{eff} was evaluated from the equilibrium temperature of the virtual blackbody probe at the ion position. The surrogate was used only to accelerate the identification; the final \textit{T}\textsubscript{eff} was evaluated with the FE model. The model yielded \textit{T}\textsubscript{eff}\textsubscript{,on}=297.6601 K, and \textit{T}\textsubscript{eff}\textsubscript{,o}\textsubscript{ff}=297.2257 K for the RF-on and RF-off conditions, respectively. The RF-induced effective-temperature rise was therefore \ensuremath{\Delta}\textit{T}\textsubscript{eff} = \textit{T}\textsubscript{eff}\textsubscript{,on} - \textit{T}\textsubscript{eff}\textsubscript{,off} =0.4345 K.

\subsection{Uncertainty-propagation framework}\label{main-sec-6.2}
The effective BBR temperature of the operational clock system is expressed as:

\begin{equation}
\begin{aligned}
T_{\mathrm{eff}}^{\mathrm{clock}}&=F\left(\hat{\mathbf{q}};\mathbf{p}_{\mathrm{fixed}},T_{\mathrm{rod}},T_{\mathrm{env}},M_{\mathrm{rad}}\right)\\
&\quad +\delta_{\mathrm{anchor}}+\delta_{\mathrm{replica}}+\delta_{\mathrm{env,tr}}+\delta_{\mathrm{spec}}
\end{aligned}
\label{main-eq-89}
\end{equation}
where \textit{F} is the coupled FE model, \ensuremath{\hat{q}} denotes the identified thermophysical parameters, p\textsubscript{fixed} collects the remaining fixed model inputs, and \textit{M}\textsubscript{rad} denotes the diffuse-grey radiation model. The additive terms account for the absolute-temperature anchoring and the transfer from the characterization system to the operational clock, where \textit{\ensuremath{\delta}}\textsubscript{anchor}, \textit{\ensuremath{\delta}}\textsubscript{replica}, \textit{\ensuremath{\delta}}\textsubscript{env,tr}, and \textit{\ensuremath{\delta}}\textsubscript{spec} denote the absolute-temperature anchoring, replica-trap transfer, environmental transfer, and spectral-emissivity transfer corrections, respectively.

For uncertainty propagation, the contributions are grouped according to the route by which they affect \textit{T}\textsubscript{eff}: (I) direct FE-model inputs, (II) inverse-identification-related contributions propagated through \ensuremath{\hat{q}}, and (III) transfer to the operational clock. Assuming the three routes to be independent, the combined standard uncertainty is

\begin{equation}
u_c^2(T_{\mathrm{eff}})=u_{\mathrm{I}}^2+u_{\mathrm{II}}^2+u_{\mathrm{III}}^2,
\label{main-eq-90}
\end{equation}
while correlations among the inverse-identified parameters are retained during their propagation rather than replaced by independent scalar uncertainties.

To avoid double counting, each uncertainty contribution is included once. The 79.0 mK chamber-surface residual of Section \ref{main-sec-5.4} is used only to validate the modelled heat-flow pathways and is therefore not directly assigned as an uncertainty of \textit{T}\textsubscript{eff}. The Pt100 calibration uncertainty is incorporated into the uncertainty of the corresponding measured thermal boundary and is not added separately. Likewise, no separate uncertainty is assigned to the surrogate field prediction because the final \textit{T}\textsubscript{eff} is calculated with the FE model; surrogate-related effects that influence the identified parameters are accounted for through Route II.

\subsection{Inverse-identification-related contributions}\label{main-sec-6.3}
\textit{Cross-condition prediction.~}The four regional standard uncertainties of the lower-voltage test agree to within 4.6\%, corresponding to a common radiance-repeatability level of approximately 0.0229 W m\textsuperscript{-}\textsuperscript{2} sr\textsuperscript{-}\textsuperscript{1}. The four regions are not repeated measurements of a single quantity: each constrains a different parameter combination, and only the black-painted region anchors the temperature-rise amplitude directly. The regional relative uncertainties are therefore averaged with equal weight, giving a repeatability-derived relative standard uncertainty of $u_{\mathrm{rel,val}}=5.711\%$. The observed prediction errors have a root mean square of 2.952\%, indicating that the prediction–measurement differences remain within the level expected from the measurement repeatability. Because the 5.711\% repeatability-derived uncertainty and the 2.952\% prediction error describe the same cross-condition discrepancy, they are not combined in quadrature; the larger value is retained as the conservative validation contribution. Since the validation compares radiance changes rather than absolute radiance levels, this relative uncertainty is propagated to the effective-temperature rise, giving

\begin{equation}
u_{\mathrm{val}}=u_{\mathrm{rel,val}}\Delta T_{\mathrm{eff}}=0.05711\times0.4345\,\mathrm{K}=24.81\,\mathrm{mK}.
\label{main-eq-92}
\end{equation}
\textit{Practical non-uniqueness.} The cross-condition test cannot reveal parameter combinations that produce nearly identical radiance distributions over the observed surfaces. Such an ambiguity remains in the multistart inversion: 23 initializations converge to a tightly clustered parameter set, while one initialization yields a distinct alternative solution with an almost indistinguishable objective value. Propagating the displacement between the main-cluster mean and the alternative solution through the FE sensitivity coefficients gives a \textit{T}\textsubscript{eff} separation of 10.6707 mK. Treating the main-cluster solution and the admissible alternative solution symmetrically as a two-point distribution gives 

\begin{equation}
u_{\mathrm{nonuniq}}=\frac{10.6707\,\mathrm{mK}}{2}=5.335\,\mathrm{mK}.
\label{main-eq-93}
\end{equation}
Because the alternative solution shifts all five identified parameters simultaneously, their effects are propagated jointly along the parameter displacement rather than combined as independent contributions. The parameter-wise decomposition of the resulting \textit{T}\textsubscript{eff} difference is given in Supplementary Table \hyperref[tabS4]{S4}. 

The image-registration contribution was evaluated by independently repeating the component-wise camera-to-surrogate registration five times. For each realization, the component-specific transformation matrices were replaced while all radiometric parameters, material masks, surrogate settings, and other experimental inputs were kept unchanged, and the complete radiance-domain inversion was repeated. The standard deviation of the resulting effective temperatures was $u_{\mathrm{reg}}=1.83\,\mathrm{mK}$, which was taken as the registration-induced standard uncertainty. The numerical-convergence contribution was evaluated from the 23 tightly clustered multistart solutions, giving $u_{\mathrm{num}}=0.2757\,\mathrm{mK}$. Both the registration and numerical-convergence contributions are assigned to Route II.

\subsection{Boundary-condition and operational-transfer contributions}\label{main-sec-6.4}
The dominant direct boundary contribution arises from the measured copper-rod boundary temperature used in the FE model. At \textit{T}\textsubscript{rod}=298.6 K, the five-point sweep in Table \hyperref[tabS7]{S7} gives a sensitivity coefficient of \textit{c}\textsubscript{rod}=0.2785 K K\textsuperscript{-1}. With the assigned \ensuremath{\pm}0.5 K rectangular interval, the corresponding standard uncertainty is

\begin{equation}
u_{\mathrm{rod}}=|c_{\mathrm{rod}}|\frac{0.5\,\mathrm{K}}{\sqrt{3}}=80.41\,\mathrm{mK}.
\label{main-eq-96}
\end{equation}
A relative standard uncertainty of 9\% is assigned to the diffuse-grey approximation used in the surface-to-surface radiation model, giving

\begin{equation}
u_{\mathrm{grey}}=0.09\Delta T_{\mathrm{eff}}=0.09\times0.4344591=39.10\,\mathrm{mK}.
\label{main-eq-97}
\end{equation}
Route III accounts for the transfer from the characterization system to the operational clock, which is housed in a different vacuum chamber. The four transfer quantities in Eq.~\eqref{main-eq-89} are included only in the uncertainty propagation and do not alter the nominal value of \textit{T}\textsubscript{eff}. 

\begin{figure}[!htbp]
\centering
\hypertarget{main-fig-14}{}
\includegraphics[width=85.0000mm,keepaspectratio]{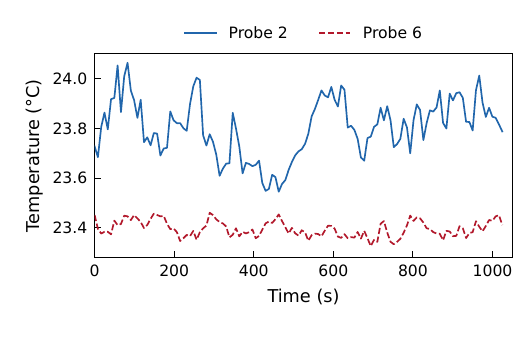}
\caption{~Temporal variation of the ambient temperature measured simultaneously by two calibrated Pt100 probes on opposite sides of the ZnSe viewport of the auxiliary characterization chamber.}\label{fig14}
\end{figure}

The absolute-temperature anchor is evaluated from the two calibrated Pt100 probes located on opposite sides of the ZnSe viewport of the auxiliary characterization chamber. In the representative 1000-s record shown in Fig. \ref{fig14}, the time-averaged temperatures of the two probes differ by 407.5 mK. The scalar auxiliary-chamber ambient temperature is represented by the mean of the two probes; therefore, half of the probe-to-probe difference, 203.8 mK, is taken as a conservative bound on the spatial representativeness of the scalar boundary temperature. Treating this bound as rectangular gives

\begin{equation}
u_{\mathrm{aux,sp}}=\frac{407.5}{2\sqrt{3}}=117.6\,\mathrm{mK}.
\label{main-eq-98}
\end{equation}
The standard uncertainty of the temporally averaged two-probe mean is 5.327 mK. Combining the spatial and temporal contributions gives

\begin{equation}
u_{\mathrm{anchor}}=\sqrt{117.6^2+5.327^2}=117.8\,\mathrm{mK}.
\label{main-eq-99}
\end{equation}
Because the probe-to-probe offset persists throughout the record, it is treated as spatial non-uniformity rather than random temporal noise and is not reduced by temporal averaging.

The environmental-transfer contribution accounts for the transfer of the thermal environment from the auxiliary characterization chamber to the operational-clock chamber. Using the ambient-temperature sensitivity coefficient \textit{c}\textsubscript{env}=0.065~K/K, together with the 175.1 mK spatial span of the seven thermometer locations forming the principal temperature band of the operational chamber, gives the conservative transfer allowance

\begin{equation}
u_{\mathrm{env,tr}}=(0.065\,\mathrm{K}/\mathrm{K})\times175.1\,\mathrm{mK}=11.38\,\mathrm{mK}.
\label{main-eq-100}
\end{equation}
The 2.34 mK calibration uncertainty of the nine-thermometer mean is substantially smaller and is not used as a substitute for the spatial-transfer allowance.

Independent long-term thermometry of the operational-clock chamber showed that temporal fluctuations were small compared with the persistent spatial temperature distribution. Across the nine retained locations, the full spatial span was 1.518 K, whereas the seven locations forming the principal temperature band spanned 175.1 mK. The complete 12-day statistics, temporal-stability analysis, and representative 100 ks record are given in Supplementary Section \hyperref[supplement-sec-S4]{S4} and Fig. \hyperlink{supplement-fig-S2}{S2}.

The remaining two transfer contributions are propagated using the FE sensitivities of \textit{T}\textsubscript{eff} to the four surface emissivities, which combine in quadrature to 111.5 mK per 100\% relative change (Supplementary Section \hyperref[supplement-sec-S2.2]{S2.2}). Assuming a 10\% relative difference between the identified in-band and total emissivities gives \textit{u}\textsubscript{spec}=0.10\ensuremath{\times}111.5=11.15 mK. Similarly, assigning a 10\% within-batch variation in the surface emissivities between the characterized and operational traps gives \textit{u}\textsubscript{replica}=11.15 mK\textit{.}

\subsection{FE-model contributions}\label{main-sec-6.5}
The nine material-property intervals listed in Supplementary Table \hyperref[tabS8]{S8}, taken as \ensuremath{\pm}10\% rectangular limits, give a combined standard uncertainty of \textit{u}\textsubscript{mat}=9.83 mK. Of this contribution, 97\% arises from the metallic thermal conductivities, while the contribution from each relative permittivity is below 0.5 mK. The individual sensitivity coefficients and uncertainty contributions are given in Supplementary Table \hyperref[tabS8]{S8}. The mesh study gives \textit{u}\textsubscript{mesh}=31.0 mK, and the geometric-tolerance sweep of S2.5 gives \textit{u}\textsubscript{geom}=3.7 mK. Because the effect of thermal-contact conductance on \textit{T}\textsubscript{eff} is nonlinear over the considered range, its uncertainty is evaluated from the full logarithmic sweep in S2.4 rather than from a local derivative. The nominal finite-contact model gives a -14.2 mK correction relative to the ideal-contact model. Over the accepted conductance range, the corresponding \textit{T}\textsubscript{eff} shift spans \ensuremath{-}29.90 to \ensuremath{-}2.21 mK. Treating this interval as rectangular gives a standard uncertainty of \textit{u}\textsubscript{contact}=8.00 mK.

\begin{equation}
\begin{aligned}
u_{\mathrm{FE,in}}&=\sqrt{u_{\mathrm{mesh}}^2+u_{\mathrm{mat}}^2+u_{\mathrm{contact}}^2+u_{\mathrm{geom}}^2}\\
&=\sqrt{31.00^2+9.83^2+8.00^2+3.70^2}\\
&=33.69\,\mathrm{mK}
\end{aligned}
\label{main-eq-102}
\end{equation}
\subsection{Combined standard uncertainty}\label{main-sec-6.6}
After all individual uncertainty contributions have been evaluated in Sections \ref{main-sec-6.3}–\ref{main-sec-6.5}, they are combined according to the three propagation routes defined in Section \ref{main-sec-6.2}. The standard uncertainties associated with Routes I–III are therefore

\begin{equation}
u_{\mathrm{I}}=\sqrt{u_{\mathrm{rod}}^2+u_{\mathrm{grey}}^2+u_{\mathrm{FE,in}}^2}=95.55\,\mathrm{mK},
\label{main-eq-103}
\end{equation}
\begin{equation}
u_{II}=\sqrt{u_{\mathrm{val}}^2+u_{\mathrm{ununiq}}^2+u_{\mathrm{reg}}^2+u_{\mathrm{num}}^2}=25.45\,\mathrm{mK},
\label{main-eq-104}
\end{equation}
\begin{equation}
u_{\mathrm{III}}=\sqrt{u_{\mathrm{anchor}}^2+u_{\mathrm{env,tr}}^2+u_{\mathrm{replica}}^2+u_{\mathrm{spec}}^2}=119.4\,\mathrm{mK}.
\label{main-eq-105}
\end{equation}
\begin{equation}
\begin{aligned}
u_c(T_{\mathrm{eff}})&=\sqrt{u_{\mathrm{I}}^2+u_{\mathrm{II}}^2+u_{\mathrm{III}}^2}\\
&=\sqrt{95.55^2+25.45^2+119.4^2}=155.0\,\mathrm{mK}.
\end{aligned}
\label{main-eq-106}
\end{equation}
This corresponds to a combined standard uncertainty of 155.0 mK. The complete uncertainty budget is summarized in Table \ref{tab8}.

\begin{table*}[!htbp]
\centering
\caption{~Uncertainty budget of the ion-position effective BBR temperature; standard uncertainties are given. The Pt100 chamber-surface comparison of Section \ref{main-sec-5.4} is a validation and is not a budget entry.}\label{tab8}
{\small\setlength{\tabcolsep}{3pt}
\begin{adjustbox}{max width=\linewidth}
\begin{tabular}{@{}llr@{}}
\toprule
Route & Contribution & \shortstack[l]{Standard uncertainty\\\textit{u}\textsubscript{i}(\textit{T}\textsubscript{eff}) / mK} \\
\midrule
I. Direct FE evaluation & Copper-rod thermal boundary & 80.41 \\
 & Diffuse-grey radiation model & 39.10 \\
 & FE-model inputs & 33.69 \\
 & Route I subtotal & 95.55 \\
\shortstack[l]{II. Inverse identification\\and cross-condition validation} & Cross-condition prediction at 485 V & 24.81 \\
 & Multistart non-uniqueness & 5.335 \\
 & Image registration & 1.83 \\
 & Numerical inversion convergence & 0.2757 \\
 & Route II subtotal & 25.45 \\
\shortstack[l]{III. Transfer to the\\operational clock} & Absolute-temperature anchor & 117.8 \\
 & Environmental transfer & 11.38 \\
 & Replica-trap transfer & 11.15 \\
 & In-band-to-total-emissivity transfer & 11.15 \\
 & Route III subtotal & 119.4 \\
Combined & Combined standard uncertainty & 155.0 \\
\bottomrule
\end{tabular}
\end{adjustbox}}
\end{table*}
\subsection{Propagation to the BBR frequency shift}\label{main-sec-6.7}
The BBR Stark shift of the E3 transition is referred to the mid-infrared anchor wavelength of Ref. \cite{m33},

\begin{equation}
\Delta\nu(T)=-\frac{\Delta\alpha(10.6\,\mu\mathrm{m})}{2h}\langle E^2\rangle_{300\mathrm{K}}\left(\frac{T}{300\,\mathrm{K}}\right)^4[1+\eta(T)],
\label{main-eq-107}
\end{equation}
with $\eta(T)$, the spectral transfer defined in Supplementary Section \hyperref[supplement-sec-S3.2]{S3.2} and$\langle E^2\rangle_{300\mathrm{K}}=(831.9\,\mathrm{V}\mathrm{m}^{-1})^2$. The reported fractional shift $y_{\mathrm{BBR}}(300\,\mathrm{K})=-81.59(24)\times10^{-18}$ \cite{m33} corresponds to $\Delta\nu(300\,\mathrm{K})=-52.39\,\mathrm{mHz}$ for $\nu_0=642.1\,\mathrm{THz}$ with the uncertainty including contributions from both the mid-infrared polarizability transfer and $\eta.$. Over the temperature range considered here, the temperature dependence is dominated by the \textit{T}\textsuperscript{4} term, giving the corresponding temperature sensitivity

\begin{equation}
S_T(T)=\frac{4|y_{\mathrm{BBR}}(300\,\mathrm{K})|}{300\,\mathrm{K}}\left(\frac{T}{300\,\mathrm{K}}\right)^3.
\label{main-eq-114}
\end{equation}
At \textit{T}\textsubscript{eff}=297.660 K, this gives \textit{S}\textsubscript{\textit{T}}=1.0626\ensuremath{\times}10\textsuperscript{-18}~K\textsuperscript{-1}, and the temperature-related uncertainty in the fractional BBR shift is

\begin{equation}
\begin{aligned}
u_T(y_{\mathrm{BBR}})&=S_T(T_{\mathrm{eff}})u(T_{\mathrm{eff}})\\
&=1.0626\times10^{-18}\,\mathrm{K}^{-1}\times0.1550\,\mathrm{K}\\
&=1.647\times10^{-19},
\end{aligned}
\label{main-eq-115}
\end{equation}
corresponding to 0.1058 mHz. The RF-induced change in the fractional BBR shift is defined as

\begin{equation}
\Delta y_{\mathrm{BBR,RF}}=y_{\mathrm{BBR}}(T_{\mathrm{eff,on}})-y_{\mathrm{BBR}}(T_{\mathrm{eff,off}}).
\label{main-eq-116}
\end{equation}
Using the fourth-power temperature dependence gives

\begin{equation}
\Delta y_{\mathrm{BBR,RF}}=-4.607\times10^{-19}.
\label{main-eq-117}
\end{equation}
This change is larger than the combined BBR uncertainty obtained below. The exact fourth-power expression and its comparison with the linear approximation are given in Supplementary Section \hyperref[supplement-sec-S3.6]{S3.6}.

The complete BBR uncertainty also includes the atomic-response uncertainty, which is taken consistently from Ref. \cite{m33}. The quoted uncertainty of 0.24\ensuremath{\times}10\textsuperscript{-18} at 300 K corresponds to a derived relative uncertainty of 0.2942\%. The atomic-response uncertainty adopted here is conditional on this transfer-based data set and does not resolve the method-dependent polarizability discrepancy discussed in Ref. \cite{m33}. Scaling this relative uncertainty to \textit{T}\textsubscript{eff}=297.660 K gives an atomic-response contribution of 2.326\ensuremath{\times}10\textsuperscript{-19}. The combined fractional uncertainty is therefore

\begin{equation}
\begin{aligned}
u_c(y_{\mathrm{BBR}})&=\sqrt{u_T^2(y_{\mathrm{BBR}})+u_{\mathrm{atom}}^2(y_{\mathrm{BBR}})}\\
&=\sqrt{1.647^2+2.326^2}\times10^{-19}\\
&=2.850\times10^{-19}
\end{aligned}
\label{main-eq-118}
\end{equation}
The atomic-response uncertainty remains the dominant contribution, while the temperature evaluation contributes approximately 33.40\% of the total variance.

\section{Discussion}\label{main-sec-7}
\begin{figure*}[!htbp]
\centering
\hypertarget{main-fig-15}{}
\includegraphics[width=165.0000mm,keepaspectratio]{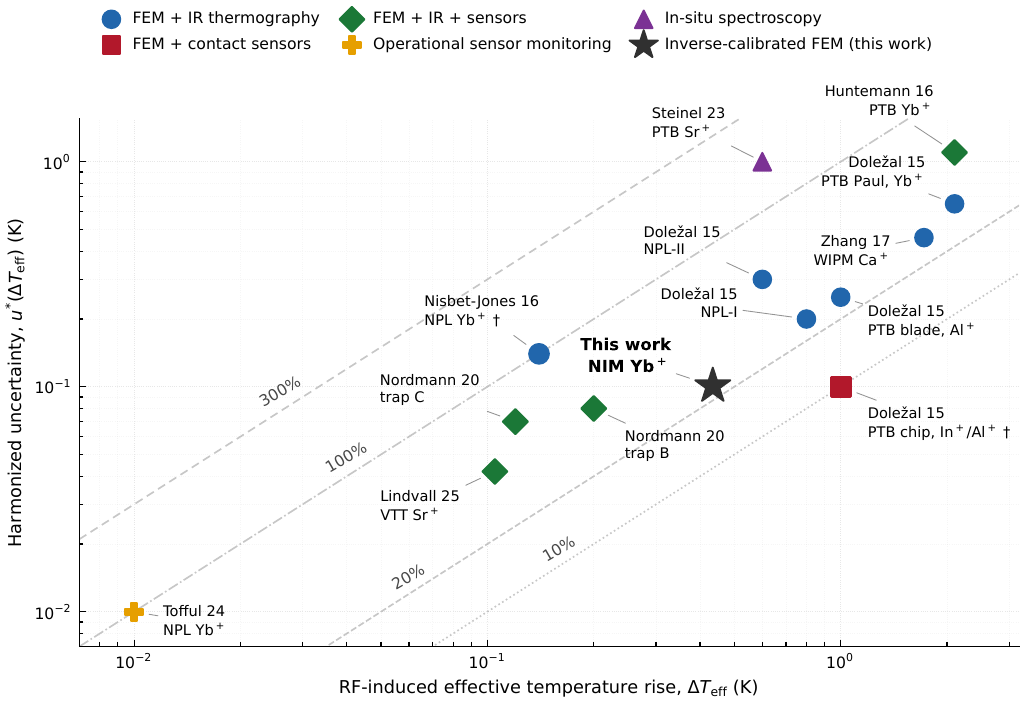}
\caption{~Harmonized effective-temperature uncertainty \textit{u}*(\ensuremath{\Delta}\textit{T}\textsubscript{eff}) against the effective-temperature rise \ensuremath{\Delta}\textit{T}\textsubscript{eff} for reported room-temperature ion-trap evaluations. Marker styles indicate the different temperature-evaluation methods. Diagonal guides indicate constant relative uncertainties \textit{u}*/\ensuremath{\Delta}\textit{T}\textsubscript{eff}. The 95\% intervals of Doležal~\textit{et al.}~\cite{m17} are converted as \textit{u}* \ensuremath{\approx} \textit{U}\textsubscript{9}\textsubscript{5}/2, whereas other values are retained as reported. Labels marked with \ensuremath{\dagger} indicate cases for which the reported uncertainty coverage is not explicit. The coverage of the Zhang~\textit{et al.}~\cite{m61} value is also unspecified. Literature values retain the source-specific budget conventions, which need not match the differential uncertainty used here. Only reported or experimentally constrained systems are shown; the model-transfer predictions of Nordmann~\textit{et al.}~\cite{m62,m31} are excluded.}\label{fig15}
\end{figure*}
The radiance-domain inversion provides an experimental constraint on thermophysical parameters that would otherwise have to be assigned mainly from literature values or nominal material properties. This is relevant for an assembled ion trap because surface condition, coating quality, and dielectric properties may differ from those of reference samples. The practical consequence is visible in the multistart analysis: an isolated alternative solution with an almost indistinguishable radiance residual differs from the main multistart solution cluster by 10.67 mK in \textit{T}\textsubscript{eff}. This shows that residual parameter non-uniqueness can already reach the 10-mK level relevant to BBR evaluation, even when the measured radiance field is reproduced well.

Figure \ref{fig15} compares the present result with reported room-temperature ion-trap evaluations. The literature data are taken from Doležal et al. \cite{m17}, Nisbet-Jones et al. \cite{m60}, Zhang et al. \cite{m61}, Huntemann et al. \cite{m32}, Nordmann et al. \cite{m62,m31}, Steinel et al. \cite{m34}, Tofful et al. \cite{m35}, and Lindvall et al. \cite{m05}. For the present work, the absolute-temperature-anchor contribution is excluded from the harmonized uncertainty because it affects the absolute temperature level rather than the differential RF-induced rise:

\begin{equation}
u^*(\Delta T_{\mathrm{eff}})=\sqrt{u_c^2(T_{\mathrm{eff}})-u_{\mathrm{anchor}}^2}=100.8\,\mathrm{mK}.
\label{main-eq-120}
\end{equation}
For \ensuremath{\Delta}\textit{T}\textsubscript{eff}=0.4345 K, this corresponds to a relative uncertainty of 23.20\%. Despite the comparatively large effective-temperature rise, the present result remains close to the 20\% guide and shows a lower relative uncertainty than several conventional FEM-based thermography evaluations at comparable or larger temperature rises. Dedicated in-situ spectroscopic approaches can achieve still lower relative uncertainties, but the present method does not require in-vacuum temperature sensors. Its main distinction is that the infrared data are used not only to observe the thermal response but also to constrain uncertain model parameters and the conductive thermal boundary.

The uncertainty budget also identifies where further improvement would be most effective. Route III accounts for 59.32\% of the total variance, with the absolute-temperature anchor alone contributing 57.74\%. The copper-rod thermal boundary is the second-largest contribution, accounting for 26.90\%, whereas the diffuse-grey radiation-model contribution accounts for 6.362\%. Further improvement should therefore focus primarily on the absolute-temperature anchoring between the characterization and operational systems, followed by improved characterization of the conductive thermal boundary.

\section{Conclusions}\label{main-sec-8}
The blackbody-radiation effective temperature of a room-temperature \textsuperscript{171}Yb\textsuperscript{+} end-cap ion trap was evaluated by combining calibrated infrared thermography with a radiance-domain inverse identification of uncertain thermophysical parameters in a finite-element model. A sensitivity-enriched reduced-order surrogate was used to accelerate the inversion, while the identified parameters were subsequently returned to the full FE model for the final \textit{T}\textsubscript{eff} evaluation. The frozen parameter set reproduces all four regional mean radiance changes within one measurement-repeatability standard uncertainty at the lower RF drive level.

The final evaluation gives \textit{T}\textsubscript{eff}=297.660 K, corresponding to an effective-temperature rise of 0.4345 K between the RF-on and RF-off conditions, with a combined standard uncertainty of 0.1550 K. This temperature uncertainty contributes 1.647\ensuremath{\times}10\textsuperscript{-19} to the fractional BBR-shift uncertainty of the \textsuperscript{2}S\textsubscript{1/2}-\textsuperscript{2}F\textsubscript{7/2} E3 transition. Combined with the atomic-response contribution, the total fractional BBR uncertainty is 2.850\ensuremath{\times}10\textsuperscript{-19}. The corresponding RF-induced change in the fractional BBR shift is \ensuremath{-}4.607\ensuremath{\times}10\textsuperscript{-19}, exceeding the combined BBR uncertainty.

The results also show why experimental identification of the thermophysical parameters is necessary. An isolated alternative inversion solution with an almost indistinguishable radiance residual differs from the main multistart solution cluster by 10.67 mK in \textit{T}\textsubscript{eff}, demonstrating that residual non-uniqueness of the model parameters can propagate to the ion-position effective temperature at the 10-mK level. The radiance-domain inversion therefore replaces poorly constrained a priori parameter values with experimentally constrained values tied to the actual assembled trap. The uncertainty analysis further shows that, once these parameters are constrained, the dominant limitation shifts from parameter uncertainty to the transfer of the thermal boundary conditions between the characterization and operational systems. This framework reduces the reliance on independent offline characterization of trap material properties and can be extended to other ion trap architectures for room-temperature optical clocks, offering a generalizable route toward further suppressing the BBR-limited uncertainty budget.

\section*{Acknowledgements}
This work was supported by the National Natural Science Foundation of China (Grant No. 12303073), the Fundamental Research Funds of the National Institute of Metrology, China (Grant No. AKYJJ2401), and the Innovation Program for Quantum Science and Technology (Grant No. 2021ZD0300901).

\section*{Data availability}
The data and code that support the findings of this study are available from the corresponding authors upon reasonable request.
\section*{Declaration of competing interest}
The authors declare that they have no known competing financial interests or personal relationships that could have appeared to influence the work reported in this paper.
\FloatBarrier

\flushcolsend
\clearpage
\raggedcolsend
\setcounter{section}{0}
\setcounter{subsection}{0}
\setcounter{subsubsection}{0}
\setcounter{figure}{0}
\setcounter{table}{0}
\setcounter{equation}{0}
\renewcommand{\theHsection}{S\arabic{section}}
\renewcommand{\theHsubsection}{\theHsection.\arabic{subsection}}
\renewcommand{\theHsubsubsection}{\theHsubsection.\arabic{subsubsection}}
\renewcommand{\theHfigure}{S\arabic{figure}}
\renewcommand{\theHtable}{S\arabic{table}}
\renewcommand{\theHequation}{S\arabic{equation}}
\renewcommand{\citenumfont}[1]{S#1}
\renewcommand{\bibnumfmt}[1]{[S#1]}
\twocolumn[{\def\baselinestretch{1}\begin{\elsarticletitlealign}
\pdfbookmark[0]{Supplementary Material}{arxiv-supplement}
\Large Supplementary Material\par\vskip6pt
\normalsize\par\vskip0pt
\footnotesize\itshape\par\vskip0pt
\hrule\vskip12pt
\end{\elsarticletitlealign}}]
\renewcommand{\thesection}{S\arabic{section}}
\renewcommand{\thefigure}{S\arabic{figure}}
\renewcommand{\thetable}{S\arabic{table}}
\renewcommand{\theequation}{S\arabic{equation}}
\section{Training and additional validation of the reduced-order surrogate}\label{supplement-sec-S1}
The architecture, reduced-order representation, parameter-to-coefficient mapping, and training objective of the surrogate are described in Section \hyperref[main-sec-3.4]{3.4} of the main text. The residual MLP was trained using the AdamW optimizer \cite{s05}. The training hyperparameters, learning-rate schedule, early-stopping criterion, and initialization settings are summarized in Table \ref{tabS1}.

\begin{table}[!htbp]
\centering
\caption{Training hyperparameters of the residual multilayer perceptron.}\label{tabS1}
{\small\setlength{\tabcolsep}{3pt}
\begin{adjustbox}{max width=\linewidth}
\begin{tabular}{@{}ll@{}}
\toprule
Setting & Value \\
\midrule
Optimizer & AdamW \\
Batch size & 256 \\
Initial learning rate & 1 \ensuremath{\times} 10\textsuperscript{-}\textsuperscript{3} \\
Weight decay & 1 \ensuremath{\times} 10\textsuperscript{-}\textsuperscript{6} \\
Sensitivity-loss weight & 0.1 \\
Maximum epochs & 1000 \\
Learning-rate scheduler & \shortstack[l]{ReduceLROnPlateau,\\factor 0.5} \\
Scheduler patience & \shortstack[l]{20 epochs without\\improvement of the\\validation criterion} \\
Minimum learning rate & 1 \ensuremath{\times} 10\textsuperscript{-}\textsuperscript{6} \\
Early stopping & \shortstack[l]{120 epochs without\\improvement} \\
Initialization seeds & 42, 2024, 3407 \\
Data-split seed & 42 \\
Selected seed / checkpoint & 2024 / epoch 290 \\
\bottomrule
\end{tabular}
\end{adjustbox}}
\end{table}
\subsection{Initialization and architecture ablation}\label{supplement-sec-S1.1}
The three initializations gave full-field validation MAEs of 0.02354, 0.02165 and 0.02338 mK for seeds 42, 2024 and 3407, with sensitivity relative RMSEs of 3.10\%, 2.45\% and 2.61\%. 

The combined validation score was defined as

\begin{equation}
S_{\mathrm{val}}=\mathrm{MAE}_{\mathrm{field}}+\lambda_{\mathrm{sel}}E_{\mathrm{sens}},
\label{supplement-eq-1}
\end{equation}
where \textit{E}\textsubscript{sens} is the dimensionless relative RMSE of the FE-sensitivity fields and $\lambda_{\mathrm{sel}}=0.05\,\mathrm{mK}$. The checkpoint with the minimum \textit{S}\textsubscript{val} was retained. Seed 2024 gave the lowest combined validation score and was adopted as the final model, at its epoch-290 checkpoint. 

The proposed model was compared with a direct CNN decoder, a conventional POD-MLP, and a quadratic-regression-only ablation using identical training, validation, and test sets, as well as the same valid-region mask. As summarized in Table \ref{tabS2}, the comparison reveals two main effects. Replacing direct pixel-wise regression with the POD representation reduced the sensitivity error by more than two orders of magnitude, from 452.5\% to 1.533\%; the quadratic baseline and the residual MLP then further reduce both the field error and the sensitivity error, with the final surrogate lowering the MAE by approximately one order of magnitude and the sensitivity error by a factor of about seven relative to the conventional POD-MLP. The residual MLP alone accounting for a 58.4\% reduction in MAE and a 74.4\% reduction in sensitivity error over the quadratic-only model. 

\begin{table*}[!htbp]
\centering
\caption{Comparison of the predictive accuracy and parameter-sensitivity fidelity of different surrogate architectures.}\label{tabS2}
{\small\setlength{\tabcolsep}{3pt}
\begin{adjustbox}{max width=\linewidth}
\begin{tabular}{@{}lrrrr@{}}
\toprule
Model & \shortstack{Test MAE\\(mK)} & \shortstack{Test RMSE\\(mK)} & \shortstack{P95 MAE\\(mK)} & \shortstack{Sensitivity\\error} \\
\midrule
Direct CNN decoder & 9.569 & 32.94 & 18.89 & 452.5\% \\
Conventional POD-MLP & 0.225 & 0.3174 & 0.6411 & 1.5330\% \\
quadratic regression only & 0.05336 & 0.07003 & 0.1357 & 0.8680\% \\
Proposed surrogate & 0.02220 & 0.03511 & 0.07432 & 0.2225\% \\
\bottomrule
\end{tabular}
\end{adjustbox}}
\end{table*}
\subsection{Additional test-set statistics}\label{supplement-sec-S1.2}
\begin{figure*}[!htbp]
\centering
\hypertarget{supplement-fig-S1}{}
\includegraphics[width=170.0000mm,keepaspectratio]{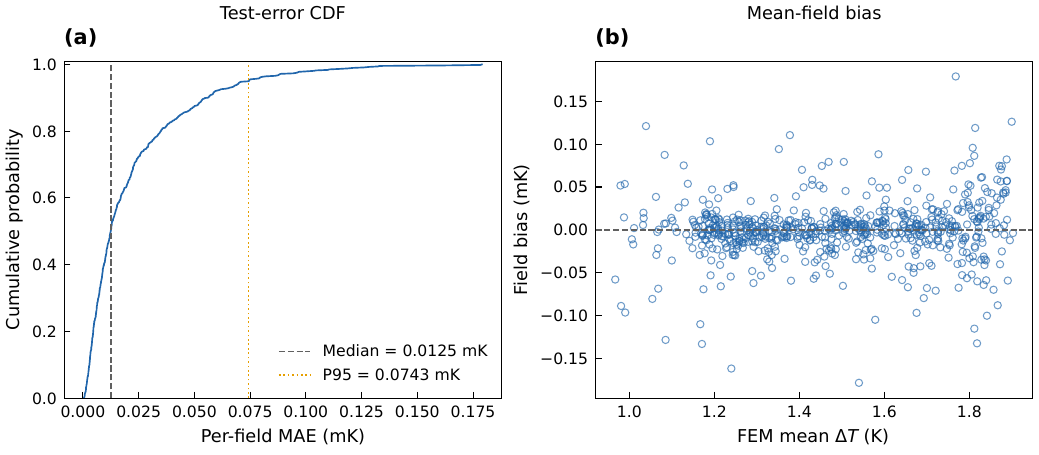}
\caption{Statistical performance of the final surrogate on the 675-case independent test set. (a) Test-error distribution of the per-field MAE; the vertical lines indicate the median value of 0.0125 mK and the 95th percentile of 0.0743 mK. (b) Mean-field prediction bias as a function of the corresponding FEM mean temperature rise.}\label{figS1}
\end{figure*}

The distribution of prediction errors over the 675-case independent test set and the dependence of the mean-field bias on the corresponding FE mean temperature rise are shown in Fig. \ref{figS1}. 

\section{Parameter and configuration sweeps}\label{supplement-sec-S2}
Supplementary Section \ref{supplement-sec-S2} provides the numerical scans supporting the multistart identifiability analysis and the FE sensitivity coefficients used in the uncertainty evaluation. Unless otherwise stated, all inputs other than the parameter under study were held at the reference values of Table \hyperref[tab5]{5}. The following sections report the complete multistart solutions, local FE sensitivity scans, copper-rod boundary sweep, thermal-contact-conductance sweep, and geometric-tolerance calculations.

\subsection{Multistart solutions of the radiance-domain inversion}\label{supplement-sec-S2.1}
The terminal parameter vectors from the 24 independent initializations used for the identifiability analysis of Section \hyperref[main-sec-4.4]{4.4} are listed in Table \ref{tabS3}. Twenty-three runs converged to a tightly clustered parameter set, whereas one run (start 19) converged to a distinct alternative solution.

\begin{table*}[!htbp]
\centering
\caption{Terminal solutions of the twenty-four multistart runs. Loss tangents in units of 10\textsuperscript{-}\textsuperscript{4} (alumina) and 10\textsuperscript{-}\textsuperscript{6} (fused silica).}\label{tabS3}
{\fontsize{8.5}{10.5}\selectfont\setlength{\tabcolsep}{3pt}
\begin{adjustbox}{max width=\linewidth}
\begin{tabular}{@{}crrrrrr@{}}
\toprule
Run & \textit{\ensuremath{\epsilon}}\textsubscript{SS} & \textit{\ensuremath{\epsilon}}\textsubscript{Cu} & \textit{\ensuremath{\epsilon}}\textsubscript{Au} & tan\textit{\ensuremath{\delta}}\textsubscript{Al}\textsubscript{\textsubscript{2}}\textsubscript{O}\textsubscript{\textsubscript{3}} & tan\textit{\ensuremath{\delta}}\textsubscript{FS} & Weighted RMS \\
\midrule
1 & 0.255574 & 0.09358455 & 0.04877349 & 1.0608 & 8.382 & 0.016058142254 \\
2 & 0.254148 & 0.09358454 & 0.04877362 & 1.0554 & 8.267 & 0.016058142261 \\
3 & 0.253181 & 0.09358453 & 0.04877371 & 1.0517 & 8.391 & 0.016058142286 \\
4 & 0.253898 & 0.09358453 & 0.04877365 & 1.0544 & 8.429 & 0.016058142265 \\
5 & 0.255111 & 0.09358454 & 0.04877351 & 1.0591 & 8.439 & 0.016058142252 \\
6 & 0.255441 & 0.09358455 & 0.04877349 & 1.0603 & 8.344 & 0.016058142253 \\
7 & 0.253273 & 0.09358453 & 0.04877372 & 1.0521 & 8.745 & 0.016058142283 \\
8 & 0.255059 & 0.09358455 & 0.04877352 & 1.0589 & 8.275 & 0.016058142252 \\
9 & 0.254920 & 0.09358455 & 0.04877355 & 1.0583 & 8.405 & 0.016058142252 \\
10 & 0.254276 & 0.09358454 & 0.04877360 & 1.0559 & 8.214 & 0.016058142259 \\
11 & 0.254969 & 0.09358455 & 0.04877354 & 1.0585 & 8.489 & 0.016058142252 \\
12 & 0.254358 & 0.09358455 & 0.04877361 & 1.0562 & 8.284 & 0.016058142257 \\
13 & 0.253556 & 0.09358455 & 0.04877370 & 1.0531 & 8.583 & 0.016058142274 \\
14 & 0.254264 & 0.09358455 & 0.04877363 & 1.0558 & 8.456 & 0.016058142258 \\
15 & 0.254751 & 0.09358454 & 0.04877356 & 1.0577 & 8.451 & 0.016058142253 \\
16 & 0.255572 & 0.09358455 & 0.04877349 & 1.0608 & 8.407 & 0.016058142254 \\
17 & 0.254815 & 0.09358454 & 0.04877352 & 1.0579 & 8.436 & 0.016058142253 \\
18 & 0.254983 & 0.09358455 & 0.04877355 & 1.0586 & 8.352 & 0.016058142252 \\
19 & 0.280856 & 0.09358470 & 0.04877203 & 1.1572 & 7.167 & 0.016058143154 \\
20 & 0.254257 & 0.09358454 & 0.04877364 & 1.0558 & 8.246 & 0.016058142259 \\
21 & 0.255035 & 0.09358454 & 0.04877354 & 1.0588 & 8.251 & 0.016058142252 \\
22 & 0.254607 & 0.09358454 & 0.04877357 & 1.0571 & 8.442 & 0.016058142254 \\
23 & 0.255155 & 0.09358456 & 0.04877356 & 1.0592 & 8.400 & 0.016058142252 \\
24 & 0.254929 & 0.09358453 & 0.04877353 & 1.0584 & 8.382 & 0.016058142252 \\
\bottomrule
\end{tabular}
\end{adjustbox}}
\end{table*}
Start 23 gave the minimum objective value and was adopted as the nominal solution. Excluding the isolated alternative solution (start 19), propagation of the 23 main-cluster terminal parameter vectors through the local FE sensitivities gives a standard deviation of 0.2757 mK in the linearized \textit{T}\textsubscript{eff}, which is used as the numerical-convergence contribution \textit{u}\textsubscript{num} in Section \hyperref[main-sec-6.3]{6.3}. Treating the main-cluster solution and the admissible alternative solution symmetrically as a two-point distribution gives \textit{u}\textsubscript{nonuniq}=10.6707/2=5.335 mK.

\begin{table*}[!htbp]
\centering
\caption{Parameter-wise decomposition of the \textit{T}\textsubscript{eff} difference between the main multistart solution cluster and the isolated alternative solution.}\label{tabS4}
{\fontsize{8.5}{10.5}\selectfont\setlength{\tabcolsep}{3pt}
\begin{adjustbox}{max width=\linewidth}
\begin{tabular}{@{}lrrrrr@{}}
\toprule
\shortstack[l]{Normalized\\parameter} & \shortstack[l]{Main-cluster\\mean} & \shortstack[l]{Alternative solution\\(start 19)} & Difference & \shortstack[l]{\ensuremath{\partial}T\textsubscript{eff}/\ensuremath{\partial}\textit{q} /\\(mK per unit q)} & \shortstack[l]{Temperature\\contribution / mK} \\
\midrule
\textit{q}(\textit{\ensuremath{\epsilon}}\textsubscript{SS}) & 0.092289 & 0.617114 & +0.524825 & 19.5877 & +10.2801 \\
\textit{q}(\textit{\ensuremath{\epsilon}}\textsubscript{Cu}) & \ensuremath{-}0.320773 & \ensuremath{-}0.320765 & +7.617\ensuremath{\times} 10\textsuperscript{-}\textsuperscript{6} & 6.28235 & +0.000048 \\
\textit{q}(\textit{\ensuremath{\epsilon}}\textsubscript{Au}) & \ensuremath{-}0.122642 & \ensuremath{-}0.122797 & \ensuremath{-}1.552\ensuremath{\times} 10\textsuperscript{-}\textsuperscript{4} & 8.61732 & \ensuremath{-}0.001337 \\
\textit{q}(tan\textit{\ensuremath{\delta}}\textsubscript{Al2O3}) & 0.574359 & 0.634696 & +0.060337 & 6.49527 & +0.39191 \\
q(tan\textit{\ensuremath{\delta}}\textsubscript{FS}) & 0.251015 & 0.158143 & \ensuremath{-}0.092871 & < 0.01 & < 0.001 \\
Linearized separation &  &  &  &  & 10.6707 \\
\bottomrule
\end{tabular}
\end{adjustbox}}
\end{table*}
\subsection{Finite-element sensitivity scans}\label{supplement-sec-S2.2}
The local sensitivities were evaluated by symmetric perturbations of \ensuremath{\Delta}q=\ensuremath{\pm}0.1 about the reference parameter set. For linearly sampled parameters, q=(x\ensuremath{-}x\textsubscript{0})/h, where x\textsubscript{0} and h are the midpoint and half-width of the interval in Table \hyperref[tab5]{5}; for the loss tangents the same normalization was applied in logarithmic coordinates. The resulting FE calculations are listed in Table \ref{tabS5}, and the corresponding central-difference sensitivities in Table \ref{tabS6}.

Table \ref{tabS7} gives the five-point copper-rod boundary sweep used in Section \hyperref[main-sec-6.4]{6.4}. A least-squares fit to all five points gives \ensuremath{\partial}T\textsubscript{eff}/\ensuremath{\partial}T\textsubscript{rod} = 0.278541 K K\textsuperscript{-}\textsuperscript{1}. The first three intervals are mutually consistent while the last is 24\% steeper, so the least-squares slope lies 4.4\% above the central difference over the inner three points, 0.26625 K K\textsuperscript{-}\textsuperscript{1}; the larger value is retained as conservative. The nominal row of this sweep and the reference row of Table \ref{tabS5} are the same configuration and differ by 4.3 mK; every sensitivity coefficient is formed from differences within a single scan, so a constant offset between scans cancels exactly.

\begin{table}[!htbp]
\centering
\caption{Raw finite-element scan about the reference parameter set. Loss tangents in units of 10\textsuperscript{-}\textsuperscript{4} (alumina) and 10\textsuperscript{-}\textsuperscript{5} (fused silica).}\label{tabS5}
{\small\setlength{\tabcolsep}{3pt}
\begin{adjustbox}{max width=\linewidth}
\begin{tabular}{@{}rrrrrrr@{}}
\toprule
$\varepsilon_{\mathrm{SS}}$ & $\varepsilon_{\mathrm{Cu}}$ & $\varepsilon_{\mathrm{Au}}$ & $\varepsilon_{\mathrm{Mo}}$ & $\tan\delta_{\mathrm{Al_2O_3}}$ & $\tan\delta_{\mathrm{FS}}$ & $T_{\mathrm{eff}} / \mathrm{K}$ \\
\midrule
0.250 & 0.1000 & 0.0500 & 0.1620 & 1.000 & 1.500 & 297.224361 \\
0.245 & 0.1000 & 0.0500 & 0.1620 & 1.000 & 1.500 & 297.222352 \\
0.255 & 0.1000 & 0.0500 & 0.1620 & 1.000 & 1.500 & 297.226269 \\
0.250 & 0.0980 & 0.0500 & 0.1620 & 1.000 & 1.500 & 297.223736 \\
0.250 & 0.1020 & 0.0500 & 0.1620 & 1.000 & 1.500 & 297.224993 \\
0.250 & 0.1000 & 0.0490 & 0.1620 & 1.000 & 1.500 & 297.223499 \\
0.250 & 0.1000 & 0.0510 & 0.1620 & 1.000 & 1.500 & 297.225223 \\
0.250 & 0.1000 & 0.0500 & 0.1588 & 1.000 & 1.500 & 297.224414 \\
0.250 & 0.1000 & 0.0500 & 0.1652 & 1.000 & 1.500 & 297.224311 \\
0.250 & 0.1000 & 0.0500 & 0.1620 & 0.861 & 1.500 & 297.223760 \\
0.250 & 0.1000 & 0.0500 & 0.1620 & 1.161 & 1.500 & 297.225059 \\
0.250 & 0.1000 & 0.0500 & 0.1620 & 1.000 & 1.270 & 297.2243609 \\
0.250 & 0.1000 & 0.0500 & 0.1620 & 1.000 & 1.780 & 297.2243611 \\
\bottomrule
\end{tabular}
\end{adjustbox}}
\end{table}
\begin{table*}[!htbp]
\centering
\caption{Closed-loop sensitivities of the ion-position effective temperature, from Table \ref{tabS5} by central differences.}\label{tabS6}
{\fontsize{8.5}{10.5}\selectfont\setlength{\tabcolsep}{3pt}
\begin{adjustbox}{max width=\linewidth}
\begin{tabular}{@{}lrrrr@{}}
\toprule
Parameter & Interval (Table \hyperref[tab5]{5}) & \shortstack[l]{\ensuremath{\partial}Teff /\ensuremath{\partial}q\\(mK per unit q)} & \ensuremath{\partial}Teff/ \ensuremath{\partial}x & \shortstack[l]{Per 100\%\\relative change / mK} \\
\midrule
\textit{\ensuremath{\epsilon}}\textsubscript{SS} & 0.20–0.30 & 19.588 & 391.8 mK & 97.94 \\
\textit{\ensuremath{\epsilon}}\textsubscript{Cu} & 0.08–0.12 & 6.282 & 314.1 mK & 31.41 \\
\textit{\ensuremath{\epsilon}}\textsubscript{Au} & 0.04–0.06 & 8.618 & 861.8 mK & 43.09 \\
\textit{\ensuremath{\epsilon}}\textsubscript{Mo} & 0.130–0.194 & \ensuremath{-}0.514 & \ensuremath{-}16.1 mK & \ensuremath{-}2.60 \\
tan\textit{\ensuremath{\delta}}\textsubscript{Al}\textsubscript{\textsubscript{2}}\textsubscript{O}\textsubscript{\textsubscript{3}} & 10\textsuperscript{-5}–2 \ensuremath{\times} 10\textsuperscript{-4} (log) & 6.495 & 4.34 mK per fold & 3.01 per doubling \\
tan\textit{\ensuremath{\delta}}\textsubscript{FS} & 10\textsuperscript{-6}–3 \ensuremath{\times} 10\textsuperscript{-5} (log) & < 0.01 & consistent with zero & < 0.01 \\
\bottomrule
\end{tabular}
\end{adjustbox}}
\end{table*}
\begin{table}[!htbp]
\centering
\caption{Five-point sweep of the copper-rod boundary temperature, all other inputs at reference values.}\label{tabS7}
{\small\setlength{\tabcolsep}{3pt}
\begin{adjustbox}{max width=\linewidth}
\begin{tabular}{@{}rrr@{}}
\toprule
$T_{\mathrm{rod}} / \mathrm{K}$ & $T_{\mathrm{eff}} / \mathrm{K}$ & \shortstack{Interval slope\\$/(\mathrm{K}\,\mathrm{K}^{-1})$} \\
\midrule
298.10 & 297.096214 & --- \\
298.35 & 297.162341 & 0.2645 \\
298.60 & 297.228680 & 0.2654 \\
298.85 & 297.295468 & 0.2671 \\
299.10 & 297.377827 & 0.3294 \\
\bottomrule
\end{tabular}
\end{adjustbox}}
\end{table}
\subsection{Thermophysical-parameter sensitivity scan}\label{supplement-sec-S2.3}
The seven thermal conductivities and two relative permittivities held fixed during the inverse identification were each varied by \ensuremath{\pm}10\% about their reference values, with all other inputs unchanged. Each interval was treated as rectangular, and the resulting local sensitivity coefficients were propagated according to the GUM. The individual contributions are listed in Table \ref{tabS8} and combine in quadrature to \textit{u}\textsubscript{mat}=9.83 mK.

The uncertainty is dominated by the metallic thermal conductivities, which account for 97\% of the total material-property variance, whereas each relative-permittivity contribution remains below 0.5 mK.

\begin{table*}[!htbp]
\centering
\caption{GUM-style propagation of the fixed FE material-property inputs. The sensitivity coefficients \textit{c}\textsubscript{i} are given in mK per displayed input unit.}\label{tabS8}
{\fontsize{8.5}{10.5}\selectfont\setlength{\tabcolsep}{3pt}
\begin{adjustbox}{max width=\linewidth}
\begin{tabular}{@{}lrrlrr@{}}
\toprule
Input quantity \textit{x}\textsubscript{i} & Estimate & \textit{u}(x\textsubscript{i}) & Evaluation & \textit{c}\textsubscript{i} & \shortstack[l]{\textit{u}\textsubscript{i}(\textit{T}\textsubscript{eff}) /\\mK} \\
\midrule
\shortstack[l]{Alumina relative\\permittivity} & 9.50 & 0.548 & Type B, rectangular & 0.903 & 0.495 \\
\shortstack[l]{Fused-silica relative\\permittivity} & 3.82 & 0.221 & Type B, rectangular & 5.68 \ensuremath{\times} 10\textsuperscript{-5} & 1.3 \ensuremath{\times} 10\textsuperscript{-5} \\
\shortstack[l]{Cu thermal conductivity\\/ W m\textsuperscript{-1} K\textsuperscript{-1}} & 400 & 23.1 & Type B, rectangular & 0.207 & 4.787 \\
\shortstack[l]{Mo thermal conductivity\\/ W m\textsuperscript{-1} K\textsuperscript{-1}} & 139 & 8.03 & Type B, rectangular & 7.80 \ensuremath{\times} 10\textsuperscript{-4} & 0.0063 \\
\shortstack[l]{Alumina thermal conductivity\\/ W m\textsuperscript{-1} K\textsuperscript{-1}} & 35.4 & 2.04 & Type B, rectangular & 0.813 & 1.662 \\
\shortstack[l]{Fused-silica thermal conductivity\\/ W m\textsuperscript{-1} K\textsuperscript{-1}} & 1.38 & 0.0797 & Type B, rectangular & 0.0232 & 0.00185 \\
\shortstack[l]{Stainless-steel thermal conductivity\\/ W m\textsuperscript{-1} K\textsuperscript{-1}} & 16.2 & 0.935 & Type B, rectangular & 8.37 & 7.824 \\
\shortstack[l]{Kovar thermal conductivity\\/ W m\textsuperscript{-1} K\textsuperscript{-1}} & 17.0 & 0.981 & Type B, rectangular & 3.13 & 3.070 \\
\shortstack[l]{ZnSe thermal conductivity\\/ W m\textsuperscript{-1} K\textsuperscript{-1}} & 18.0 & 1.04 & Type B, rectangular & 0.177 & 0.184 \\
Material-property subtotal &  &  & RSS &  & 9.83 \\
\bottomrule
\end{tabular}
\end{adjustbox}}
\end{table*}
\subsection{Thermal-contact-conductance sweep}\label{supplement-sec-S2.4}
Three bolted interfaces were assigned finite thermal contact conductance: feedthrough–stainless steel (hc\textsubscript{1}), feedthrough–copper (hc\textsubscript{2}) and copper–fused-silica/end-cap electrode (hc\textsubscript{3}). If treated using the same local \ensuremath{\pm}10\% perturbation adopted for the fixed thermophysical parameters, their combined standard uncertainty is only 0.508 mK and is dominated by hc\textsubscript{1} (Table \ref{tabS9}). 

This local estimate, however, substantially understates the actual contribution. The thermal contact conductance of a bolted joint is not known within a narrow relative interval because it depends on surface finish, bolt preload and assembly history. Over the relevant conductance range, the local derivative varies strongly, so the sensitivity evaluated at the nominal point is not representative.  The uncertainty contribution used in Section \hyperref[main-sec-6.5]{6.5} was therefore derived from the complete conductance sweep instead. Relative to idealized perfect contact, the nominal 1 \ensuremath{\times} 10\textsuperscript{3} W m\textsuperscript{-}\textsuperscript{2} K\textsuperscript{-}\textsuperscript{1} shifts \textit{T}\textsubscript{eff} by \ensuremath{-}14.2 mK. Over 2 \ensuremath{\times} 10\textsuperscript{2} to 1 \ensuremath{\times} 10\textsuperscript{4} W m\textsuperscript{-}\textsuperscript{2} K\textsuperscript{-}\textsuperscript{1}, the shift spans \ensuremath{-}29.9 to \ensuremath{-}2.2 mK; treating this interval as rectangular gives \textit{u}\textsubscript{contact}=8.00 mK.

\begin{table*}[!htbp]
\centering
\caption{Local sensitivity of \textit{T}\textsubscript{eff} to the three thermal contact conductances at the nominal 1 \ensuremath{\times} 10\textsuperscript{3} W m\textsuperscript{-}\textsuperscript{2} K\textsuperscript{-}\textsuperscript{1}. }\label{tabS9}
{\small\setlength{\tabcolsep}{3pt}
\begin{adjustbox}{max width=\linewidth}
\begin{tabular}{@{}lrrr@{}}
\toprule
Interface & \shortstack{$c_i / (\mathrm{K}/$\\$(\mathrm{W}\,\mathrm{m}^{-2}\,\mathrm{K}^{-1}))$} & \shortstack{$\pm10\%$ excursion\\/ mK} & $u_i / \mathrm{mK}$ \\
\midrule
$\mathrm{hc}_{1}$ (feedthrough--stainless steel) & $8.80\times10^{-6}$ & 0.880 & 0.508 \\
$\mathrm{hc}_{3}$ (copper--electrode) & $-1.17\times10^{-7}$ & 0.0117 & 0.0068 \\
$\mathrm{hc}_{2}$ (feedthrough--copper) & $-6.82\times10^{-10}$ & $6.8\times10^{-5}$ & $3.9\times10^{-5}$ \\
Combined &  &  & 0.508 \\
\bottomrule
\end{tabular}
\end{adjustbox}}
\end{table*}
\begin{table}[!htbp]
\centering
\caption{Effect of the assumed thermal contact conductance on \textit{T}\textsubscript{eff}, relative to idealized perfect contact.}\label{tabS10}
{\small\setlength{\tabcolsep}{3pt}
\begin{adjustbox}{max width=\linewidth}
\begin{tabular}{@{}lr@{}}
\toprule
hc / (W m\textsuperscript{-}\textsuperscript{2} K\textsuperscript{-}\textsuperscript{1}) & \textit{T}\textsubscript{eff} shift / mK \\
\midrule
2 \ensuremath{\times} 10\textsuperscript{2} & -29.90 \\
3 \ensuremath{\times} 10\textsuperscript{2} & \ensuremath{-}26.51 \\
1 \ensuremath{\times} 10\textsuperscript{3} (nominal) & \ensuremath{-}14.16 \\
3 \ensuremath{\times} 10\textsuperscript{3} & \ensuremath{-}6.25 \\
1 \ensuremath{\times} 10\textsuperscript{4} & \ensuremath{-}2.21 \\
\bottomrule
\end{tabular}
\end{adjustbox}}
\end{table}
\subsection{Geometric-tolerance sweep}\label{supplement-sec-S2.5}
The seven FE configurations used to evaluate the geometric-tolerance contribution reported in Sections \hyperref[main-sec-3.2]{3.2} and \hyperref[main-sec-6.5]{6.5} of the main text are listed in Table \ref{tabS11}. The configurations comprise the nominal geometry, \ensuremath{\pm}0.05 mm variations of the left–right end-cap separation and the left–right Mo RF-electrode separation, and \ensuremath{\pm}5\ensuremath{{}^{\circ}} relative angular misalignment between the two Mo RF electrodes.

\begin{table*}[!htbp]
\centering
\caption{\textit{T}\textsubscript{eff} for seven electrode configurations spanning the assembly tolerances. Displacements per electrode.}\label{tabS11}
{\fontsize{8.5}{10.5}\selectfont\setlength{\tabcolsep}{3pt}
\begin{adjustbox}{max width=\linewidth}
\begin{tabular}{@{}lrrrrrrrr@{}}
\toprule
Configuration & y\textsubscript{1} / mm & y\textsubscript{2} / mm & y\textsubscript{3} / mm & y\textsubscript{4} / mm & \ensuremath{\theta} / \ensuremath{{}^{\circ}} & \textit{T}\textsubscript{eff} / K & \ensuremath{\Delta}\textit{T}\textsubscript{eff} / K & \shortstack[l]{\textit{T}\textsubscript{eff} - \textit{T}\textsubscript{norm} /\\mK} \\
\midrule
Nominal & 0 & 0 & 0 & 0 & 0 & 297.2286 & 0.5386 & 0 \\
Cap-end spacing \ensuremath{-}0.05 mm & +0.025 & \ensuremath{-}0.025 & 0 & 0 & 0 & 297.2318 & 0.5418 & +3.2 \\
Cap-end spacing +0.05 mm & \ensuremath{-}0.025 & +0.025 & 0 & 0 & 0 & 297.2259 & 0.5359 & \ensuremath{-}2.7 \\
\shortstack[l]{Mo RF-electrode\\spacing 0.95 mm} & 0 & 0 & +0.025 & \ensuremath{-}0.025 & 0 & 297.2311 & 0.5411 & +2.5 \\
\shortstack[l]{Mo RF-electrode\\spacing 1.05 mm} & 0 & 0 & \ensuremath{-}0.025 & +0.025 & 0 & 297.2300 & 0.5400 & +1.4 \\
\shortstack[l]{Mo RF-electrode\\misalignment +5\ensuremath{{}^{\circ}}} & 0 & 0 & 0 & 0 & +5 & 297.2362 & 0.5462 & +7.6 \\
\shortstack[l]{Mo RF-electrode\\misalignment \ensuremath{-}5\ensuremath{{}^{\circ}}} & 0 & 0 & 0 & 0 & \ensuremath{-}5 & 297.2359 & 0.5459 & +7.3 \\
\bottomrule
\end{tabular}
\end{adjustbox}}
\end{table*}
\section{Atomic-response model and BBR-shift propagation}\label{supplement-sec-S3}
This section summarizes the atomic-response relations and spectral-transfer treatment used to convert the effective BBR temperature evaluated in the main text into the fractional BBR shift of the \textsuperscript{1}\textsuperscript{7}\textsuperscript{1}Yb\textsuperscript{+} \textsuperscript{2}S\textsubscript{1}/\textsubscript{2} (F = 0)–\textsuperscript{2}F\textsubscript{7}/\textsubscript{2} (F = 3) E3 clock transition. The thermal evaluation developed in this work is independent of the adopted atomic-response data set. The latter is therefore treated separately, and central values and uncertainties are taken consistently from a single atomic-response set rather than mixed between different determinations.

\subsection{Effective BBR temperature and field equivalence}\label{supplement-sec-S3.1}
For a non-isothermal radiative environment, the ion-position effective temperature is defined through the total mean-square electric field,

\begin{equation}
\langle E^2\rangle_{\mathrm{ion}}\equiv\langle E^2\rangle_{300}\left(\frac{T_{\mathrm{eff}}}{300\,\mathrm K}\right)^4
\label{supplement-eq-8}
\end{equation}
The blackbody sphere of Section \hyperref[main-sec-2.1]{2.1} realizes this definition: a small blackbody in radiative equilibrium settles at the temperature of an isotropic field of the same \ensuremath{\langle}E\textsuperscript{2}\ensuremath{\rangle}. The Planck spectral energy density per unit angular frequency \cite{s14} is

\begin{equation}
u_{\omega}(T)=\frac{\hbar\omega^3}{\pi^2c^3}\frac{1}{\exp\!\left(\frac{\hbar\omega}{k_{B}T}\right)-1},
\label{supplement-eq-9}
\end{equation}
and its integration gives

\begin{equation}
\langle E^2(T)\rangle=\frac{\pi^2(k_{B}T)^4}{15\varepsilon_0\hbar^3c^3}=(831.943\,\mathrm{V\,m}^{-1})^2\left(\frac{T}{300\,\mathrm K}\right)^4.
\label{supplement-eq-10}
\end{equation}
\subsection{E3 BBR shift and spectral transfer}\label{supplement-sec-S3.2}
The scalar differential polarizability of the two clock states is

\begin{equation}
\Delta\alpha(\omega)=\alpha_e^{(0)}(\omega)-\alpha_g^{(0)}(\omega),\qquad\Delta\alpha_{\mathrm{dc}}=\Delta\alpha(0)
\label{supplement-eq-11}
\end{equation}
and the corresponding static electric-dipole BBR Stark shift is

\begin{equation}
\Delta\nu_{\mathrm{st}}(T)=-\frac{\Delta\alpha_{\mathrm{dc}}}{2h}\langle E^2\rangle_{300}\left(\frac{T}{300\,\mathrm K}\right)^4.
\label{supplement-eq-12}
\end{equation}
The full Planck-weighted shift can be written in a dc-anchored form \cite{s14,s15}, 

\begin{equation}
\Delta\nu_{\mathrm{BBR}}(T)=-\frac{\Delta\alpha_{\mathrm{dc}}}{2h}\langle E^2\rangle_{300}\left(\frac{T}{300\,\mathrm K}\right)^4[1+\eta(T)],
\label{supplement-eq-13}
\end{equation}
whose dynamic correction \textit{\ensuremath{\eta}} is defined by a spectral integral over the Planck distribution,

\begin{equation}
1+\eta(T)=\frac{\displaystyle\int_0^{\infty}\Delta\alpha(\omega)\frac{d\langle E^2\rangle_T}{d\omega}\,d\omega}{\Delta\alpha(0)\langle E^2(T)\rangle},
\label{supplement-eq-14}
\end{equation}
where $\frac{d\langle E^2\rangle_T}{d\omega}=\frac{\hbar\omega^3}{\pi^2\varepsilon_0c^3}\frac{1}{\exp\!\left(\frac{\hbar\omega}{k_BT}\right)-1}$.

If the experimental reference is instead a mid-infrared polarizability \ensuremath{\Delta}\textit{\ensuremath{\alpha}}(\textit{\ensuremath{\lambda}}\textsubscript{ref}), the corresponding correction is

\begin{equation}
\omega_r=\frac{2\pi c}{\lambda_{\mathrm{ref}}},
\label{supplement-eq-16}
\end{equation}
\begin{equation}
1+\xi(T;\omega_r)=\frac{\displaystyle\int_0^{\infty}\Delta\alpha(\omega)\frac{d\langle E^2\rangle_T}{d\omega}\,d\omega}{\Delta\alpha(\omega_r)\langle E^2(T)\rangle}.
\label{supplement-eq-17}
\end{equation}
The two forms have different denominators and cannot be exchanged without changing the central reference polarizability. Differentiating with respect to temperature,

\begin{equation}
y_{\mathrm{BBR}}(T)=C\left(\frac{T}{300\,\mathrm K}\right)^4[1+\eta(T)],
\label{supplement-eq-18}
\end{equation}
where $C=-\frac{\Delta\alpha_{\mathrm{dc}}\langle E^2\rangle_{300}}{2h\nu_0}$,

then $\frac{\partial y_{\mathrm{BBR}}}{\partial T}=C\left[\frac{4}{300\,\mathrm K}\left(\frac{T}{300\,\mathrm K}\right)^3[1+\eta(T)]+\left(\frac{T}{300\,\mathrm K}\right)^4\frac{d\eta}{dT}\right]$,

\begin{equation}
\frac{\partial y_{\mathrm{BBR}}}{\partial T}=y_{\mathrm{BBR}}(T)\left[\frac{4}{T}+\frac{1}{1+\eta(T)}\frac{d\eta}{dT}\right],
\label{supplement-eq-21}
\end{equation}
the shift scales predominantly as T\textsuperscript{4} and its temperature sensitivity predominantly as T\textsuperscript{3}.

\subsection{Experimental determinations of the differential polarizability}\label{supplement-sec-S3.3}
The scalar electric-dipole polarizability of a clock state is a sum over states,

\begin{equation}
\begin{aligned}
\alpha_a^{(0)}(\omega)&=\frac{2}{3(2J_a+1)}\sum_n\frac{\Delta E_{na}\left|\langle n\|D\|a\rangle\right|^2}{(\Delta E_{na})^2-(\hbar\omega)^2}\\
&\quad +\alpha_{\mathrm{core}}+\alpha_{\mathrm{vc}}+\alpha_{\mathrm{tail}}
\end{aligned}
\label{supplement-eq-22}
\end{equation}
in which measured lifetimes and branching fractions can replace theoretical line strengths, the lifetime-assisted route used in many clock evaluations \cite{s06,s07}. For the Yb\textsuperscript{+} E3 transition this route is not sufficient on its own: the upper f-hole state has strong configuration mixing, so complete branching information, core and tail terms and correlation corrections are all required \cite{s08,s09}. A lifetime table alone therefore determines neither \ensuremath{\Delta}\textit{\ensuremath{\alpha}} nor the spectral curvature entering \textit{\ensuremath{\eta}}.

Four experimental routes are relevant.

\textit{Huntemann} \textit{et al.~\cite{s15}} measured scalar shifts at 852, 1064, 1310 and 1554 nm and extrapolated to dc. This peer-reviewed dc-anchored set is the principal alternative to the set adopted here.

\textit{Baynham et al.~\cite{s10}} measured at 7.17 µm, close to the room-temperature BBR spectrum, and extrapolated to dc. It is a useful spectral cross-check, but the authors state that unmeasured beam-profile power prevents full validation of the quoted uncertainty.

\textit{Barrett et al.~\cite{s11,s12}} combine reference-ion intensity calibration, measured polarizability ratios and atomic-structure information while retaining the wavelength dependence and its covariance, providing a framework for combining sparse measurements with theory.

\textit{Steinel \cite{s16}} used an \textsuperscript{8}\textsuperscript{8}Sr\textsuperscript{+}-referenced mid-infrared transfer. This is the most precise reported route, but its central value lies about 13\% above the Huntemann/Baynham scale when the two are compared as 300 K fractional shifts; the discrepancy is unresolved.

Frequency-dependent polarizabilities must not be relabelled as dc values without an explicit spectral model. All polarizabilities here are in SI units; with 1 a.u. = 1.64877 \ensuremath{\times} 10\textsuperscript{-}\textsuperscript{4}\textsuperscript{1} J m\textsuperscript{2} V\textsuperscript{-}\textsuperscript{2}, the Huntemann and Steinel values correspond to 5.39 and 6.08 a.u. respectively.

\subsection{Coherent atomic-response sets}\label{supplement-sec-S3.4}
The central BBR correction and the atomic-response uncertainty must be taken from the same data set. Three internally coherent choices illustrate the present state of the literature: the Huntemann dc set \cite{s15}, the Steinel rf-transfer set \cite{s16} adopted here, and the Steinel light-shift-referenced set \cite{s16}.

These are alternative data sets, not repeated measurements to be averaged. In particular the Huntemann central polarizability must not be combined with the Steinel rf-transfer uncertainty, which would mix different central values with different dynamic-correction normalizations. The rf-transfer set is used throughout the present manuscript for two reasons: it is the only route whose reported uncertainty is comparable to the temperature-related term evaluated here, and its mid-infrared anchor lies close to the spectral centroid of the room-temperature blackbody field, so the transfer to the blackbody spectrum is small. If the discrepancy between the routes later be resolved in favor of the dc-anchored scale, both the atomic-response term and the sensitivity coefficient must be rescaled in proportion to the adopted central shift.

\subsection{GUM formulation and consistency check}\label{supplement-sec-S3.5}
Following the GUM \cite{s13}, the fractional BBR shift is treated as a function of the thermal and atomic-response input quantities,

\begin{equation}
y_{\mathrm{BBR}}=f(T_{\mathrm{eff}},\Delta\alpha,\eta,\nu_0),
\label{supplement-eq-23}
\end{equation}
\begin{equation}
\mathbf{x}=(T_{\mathrm{eff}},\Delta\alpha,\eta,\nu_0)^{\mathrm T},
\label{supplement-eq-24}
\end{equation}
\begin{equation}
u^2(y_{\mathrm{BBR}})=\mathbf J\mathbf U_x\mathbf J^{\mathrm T},
\label{supplement-eq-25}
\end{equation}
with the corresponding sensitivity coefficients

\begin{equation}
\mathbf J=\left(\frac{\partial y_{\mathrm{BBR}}}{\partial T},\frac{y_{\mathrm{BBR}}}{\Delta\alpha(\omega_r)},\frac{y_{\mathrm{BBR}}}{1+\xi},-\frac{y_{\mathrm{BBR}}}{\nu_0}\right).
\label{supplement-eq-26}
\end{equation}
The thermal evaluation of this work is independent of the selected atomic-response set, so the thermal and atomic-response contributions are combined in quadrature,

\begin{equation}
\begin{aligned}
u^2(y_{\mathrm{BBR}})&=u_T^2(y_{\mathrm{BBR}})+u_{\mathrm{atom}}^2(y_{\mathrm{BBR}}),\\
u_T(y_{\mathrm{BBR}})&=S_Tu_c(T_{\mathrm{eff}}),
\end{aligned}
\label{supplement-eq-27}
\end{equation}
\begin{equation}
S_T=\left|\frac{\partial y_{\mathrm{BBR}}}{\partial T}\right|_{T=T_{\mathrm{eff}}}.
\label{supplement-eq-28}
\end{equation}
Evaluating the adopted rf-transfer set at the reference wavelength 10.6 µm reproduces the 300 K fractional shift reported in Ref. \cite{s16} to within 0.02\%, which checks the whole propagation chain. The numerical propagation at the present operating point is reported in Section \hyperref[main-sec-6.7]{6.7} of the main text.

\subsection{Exact RF-induced BBR correction}\label{supplement-sec-S3.6}
The BBR correction associated specifically with the RF-induced effective-temperature rise is evaluated from the exact fourth-power temperature dependence,

\begin{equation}
\delta y_{\mathrm{RF}}=y_{\mathrm{BBR}}(300\,\mathrm K)\left[\left(\frac{T_{\mathrm{eff,on}}}{300\,\mathrm K}\right)^4-\left(\frac{T_{\mathrm{env,off}}}{300\,\mathrm K}\right)^4\right].
\label{supplement-eq-29}
\end{equation}
The linear approximation \textit{S}\textsubscript{\textit{T}}\ensuremath{\Delta}\textit{T}\textsubscript{eff} differs from the exact fourth-power expression by approximately 0.22\%; the exact form is therefore used in Section \hyperref[main-sec-6.7]{6.7}.

The temperature-related uncertainty and the atomic-response uncertainty are reported separately so that the thermal result of the present work can be recombined with an alternative internally consistent atomic-response data set if future measurements resolve the discrepancy between the available determinations.

\section{Thermal environment of the operational-clock chamber}\label{supplement-sec-S4}
The thermophysical-parameter identification described in Sections \hyperref[main-sec-4]{4} and \hyperref[main-sec-5]{5} was performed using the trap installed in the auxiliary characterization chamber, whereas the operational clock is housed in the main chamber inside the magnetic shield. These two thermal environments are treated separately in the uncertainty evaluation. The two calibrated Pt100 probes located on opposite sides of the ZnSe viewport of the auxiliary chamber provide the ambient-temperature anchor used during thermographic characterization and are discussed in Section \hyperref[main-sec-6.4]{6.4} of the main text. The present section characterizes the thermal environment of the operational-clock chamber and provides the experimental basis for the environmental-transfer contribution.

Ten calibrated Pt100 thermometers were distributed over the outer wall of the main chamber and logged at 10 s intervals from 2026-02-12 to 2026-03-02, corresponding to 17.8 days and 153825 samples with a single 56 s interruption. Two intervals were excluded: the first day, during which the chamber relaxed by approximately 2 K after pump-down, and a disturbance on 2026-02-28 during which one channel exhibited a 1.4 K excursion. The remaining audited interval extends from 2026-02-14 to 2026-02-26, corresponding to 12.0 days. Channel 8 was excluded from the following analysis because its temporal behaviour was inconsistent with that of the remaining sensors and was attributed to an instrumental artefact.

\subsection{Spatial structure and its origin}\label{supplement-sec-S4.1}
Over the audited 12-day interval, the nine retained thermometer locations had an overall mean temperature of 299.8433 K and a full spatial span of 1.518 K. Seven locations formed a principal temperature band spanning 175.1 mK, with a standard deviation across these locations of approximately 66 mK. The remaining two locations lay approximately 0.72 K above and 0.80 K below the nine-channel mean, respectively.

These persistent offsets are interpreted as a spatial temperature distribution of the chamber rather than as random thermometer noise. The pattern is consistent with the dominant conductive heat-flow path from the helical-resonator/copper-rod side of the apparatus into the chamber structure. Consequently, the thermometer array is used to characterize the spatial structure of the operational-clock thermal environment rather than as nine repeated measurements of a single uniform temperature.

Where a scalar clock-side environmental temperature is required for reporting purposes, the mean of the nine retained thermometers is used. This scalar quantity does not replace the auxiliary-chamber ambient temperature used during parameter identification; instead, the spatial distribution measured on the operational chamber provides the experimental basis for the environmental-transfer allowance introduced in Section \hyperref[main-sec-6.4]{6.4} of the main text.

\begin{table}[!htbp]
\centering
\caption{Per-channel temperature statistics of the operational-clock chamber. The first three statistical columns refer to the audited 12.0-day interval, whereas the final three refer to the 100 ks interval shown in Fig. \ref{figS2}. Channel 8 was excluded because of instrumental instability.}\label{tabS12}
{\small\setlength{\tabcolsep}{3pt}
\begin{adjustbox}{max width=\linewidth}
\begin{tabular}{@{}crrrrrr@{}}
\toprule
Channel & \shortstack{12 d mean\\/ K} & \shortstack{12 d sd\\/ mK} & \shortstack{12 d ptp\\/ mK} & \shortstack{100 ks\\sd / mK} & \shortstack{100 ks\\ptp / mK} & \shortstack{100 ks\\drift / mK} \\
\midrule
1 & 300.5607 & 41.0 & 147.0 & 2.3 & 11.3 & $+0.6$ \\
2 & 299.0429 & 36.8 & 127.2 & 2.6 & 13.9 & $-0.4$ \\
3 & 299.8733 & 40.0 & 138.7 & 2.3 & 10.6 & $+0.7$ \\
4 & 299.7800 & 39.9 & 137.8 & 2.3 & 10.6 & $+0.7$ \\
5 & 299.7671 & 40.4 & 139.7 & 2.2 & 10.3 & $+0.7$ \\
6 & 299.8339 & 40.2 & 138.6 & 2.2 & 10.3 & $+0.7$ \\
7 & 299.8719 & 42.5 & 148.4 & 2.3 & 10.6 & $+0.6$ \\
9 & 299.9180 & 39.6 & 136.5 & 2.3 & 10.8 & $+0.7$ \\
10 & 299.9422 & 40.7 & 143.0 & 2.3 & 11.6 & $+0.5$ \\
Mean & 299.8433 &  &  &  &  &  \\
\bottomrule
\end{tabular}
\end{adjustbox}}
\end{table}
\subsection{Temporal stability and the uncertainty of Tenv}\label{supplement-sec-S4.2}
\begin{figure*}[!htbp]
\centering
\hypertarget{supplement-fig-S2}{}
\includegraphics[width=120.9448mm,keepaspectratio]{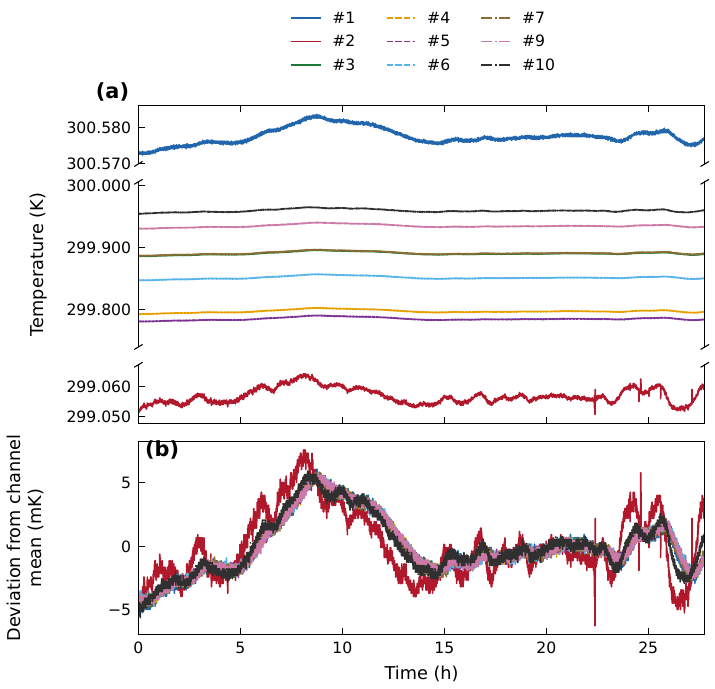}
\caption{Thermal record of the operational-clock chamber over the 100 ks interval beginning on 2026-02-23 at 09:59:56. (a) Absolute temperatures of the nine retained Pt100 thermometers, with the two extreme locations displayed on separate broken-axis ranges. (b) The same records after subtraction of each channel's own temporal mean, showing predominantly common-mode temporal variation while the persistent spatial offsets between thermometer locations remain.}\label{figS2}
\end{figure*}

The displayed interval of Fig. \ref{figS2} is the 100 ks (27.8 h) beginning 2026-02-23 09:59:56. It was selected within the audited span as the gap-free, step-free interval whose worst-channel drift and channel-to-channel tracking are jointly smallest; 521 gap-free candidates were scored at 1800 s spacing. Shorter intervals were rejected for display: the best 1 ks and 10 ks candidates vary by 1.0 and 1.8 mK peak to peak, four to seven quantization steps of the logger (0.258 mK), and therefore show the digitization floor rather than the environment.

Within the displayed interval each retained channel has a standard deviation of 2.2–2.6 mK and a peak-to-peak excursion of 10.3–13.9 mK, with an end-to-end drift of at most 0.7 mK. Referred to each channel’s own mean the nine records agree to 1.0 mK root mean square, with a worst instantaneous spread of 6.3 mK, against a common-mode excursion of 10.2 mK: the temporal variation is predominantly common-mode while the spatial offsets remain persistent.

\begin{table*}[!htbp]
\centering
\caption{Overlapping Allan deviation of the nine-channel mean over the audited span.}\label{tabS13}
{\fontsize{8.5}{10.5}\selectfont\setlength{\tabcolsep}{3pt}
\begin{adjustbox}{max width=\linewidth}
\begin{tabular}{@{}lrrrrrrrrr@{}}
\toprule
\ensuremath{\tau} / s & 10 & 30 & 100 & 300 & 1000 & 3000 & 10 000 & 30 000 & 100 000 \\
\midrule
\textit{\ensuremath{\sigma}}\textsubscript{\textit{T}} / mK & 0.092 & 0.083 & 0.103 & 0.235 & 0.725 & 2.00 & 5.36 & 10.47 & 19.18 \\
\bottomrule
\end{tabular}
\end{adjustbox}}
\end{table*}
The deviation reaches 0.083 mK at \textit{\ensuremath{\tau}} = 30 s and rises approximately in proportion to \textit{\ensuremath{\tau}} beyond a few hundred seconds, the signature of drift rather than of white noise; averaging therefore gains nothing past about one minute. Over a thermogram acquisition the Type A contribution to the nine-channel mean is consequently below 0.11 mK and is negligible.

The uncertainty of \textit{T}\textsubscript{env} is then set by the calibration. Of the four terms of Table \hyperref[tab1]{1}, the fitting residual \textit{u}\textsubscript{fit} = 2.02 mK is a property of the individual channel and averages down over the n = 9 retained thermometers, whereas the SPRT calibration \textit{u}\textsubscript{SPRT} = 2 mK, the bath inhomogeneity \textit{u}\textsubscript{bath} = 1 mK and the bath stability \textit{u}\textsubscript{stab} = 0.16 mK are common to all channels. This gives

\begin{equation}
\begin{aligned}
u(T_{\mathrm{env,main}})&=\sqrt{\frac{u_{\mathrm{fit}}^2}{n}+u_{\mathrm{SPRT}}^2+u_{\mathrm{bath}}^2+u_{\mathrm{stab}}^2}\\
&=\sqrt{\left(\frac{2.02}{3}\right)^2+2.00^2+1.00^2+0.16^2}\\
&=2.34\,\mathrm{mK},
\end{aligned}
\label{supplement-eq-31}
\end{equation}
against 3.01 mK for a single worst-case channel. It is much smaller than the persistent spatial variation of the operational chamber and is therefore not used as the environmental-transfer term itself. Instead, the transfer allowance is based on the experimentally observed clock-side spatial temperature distribution, as described in Section \hyperref[main-sec-6.4]{6.4} of the main text.

\section{Traceability of the numerical evidence}\label{supplement-sec-S5}
The terminal solutions of the twenty-four multistart runs (Table \ref{tabS3}), together with the parameter-displacement propagation in Table \ref{tabS4}, support the multistart non-uniqueness allowance and the numerical-convergence contribution of Section \hyperref[main-sec-6.3]{6.3}. The finite-element sensitivity scans (Tables \ref{tabS5}–\ref{tabS7}) and the thermal-contact-conductance sweeps (Tables \ref{tabS9}–\ref{tabS10}) support, respectively, the closed-loop sensitivity coefficients of Sections \hyperref[main-sec-6.4]{6.4} and \hyperref[main-sec-6.5]{6.5} and the contact-conductance contribution to the FE input-propagation term. The geometric-tolerance sweep (Table \ref{tabS11}) supports the \(u_{\mathrm{geom}}\) entry of Section \hyperref[main-sec-6.5]{6.5}. Section \ref{supplement-sec-S3} documents the atomic-response relations and the coherent data set used to convert the thermal result into a fractional BBR shift. Section \ref{supplement-sec-S4} reports the independent 12-day operational-clock chamber thermometry (Tables \ref{tabS12} and \ref{tabS13} and Fig. \ref{figS2}), which provides the experimental basis for the environmental-transfer contribution of Section \hyperref[main-sec-6.4]{6.4}.

\FloatBarrier

\end{document}